\documentclass{article}
\usepackage{amssymb}
\usepackage{fullpage}
\usepackage{amsmath}
\usepackage{amsfonts}
\usepackage{graphicx}
\usepackage{float}
\begin{document}

\title{On The Existence Of Anisotropic Cosmological Models In
Higher-Order Theories Of Gravity}
\author{Jonathan Middleton \\
DAMTP, Centre for Mathematical Sciences, \\
Cambridge University, \\
Wilberforce Rd., Cambridge CB3 0WA, UK}
\maketitle

\abstract{We investigate the behaviour on approach to the initial singularity in higher-order extensions of general relativity by finding exact cosmological solutions for a wide class of models in which the Lagrangian is allowed to depend nonlinearly upon the three possible linear and quadratic scalars built from the Riemann tensor; $R$, $R_{ab}R^{ab}$ and $R_{abcd}R^{abcd}$. We present new anisotropic vacuum solutions analagous to the Kasner solutions of general relativity and extend previous results to a much wider range of fourth order theories of gravity. We discuss the implications of these results for the behaviour of the more general anisotropic Bianchi type VIII and IX cosmologies as the initial singularity is approached %and find that the infinite sequence of chaotic Mixmaster oscillations between different Kasner regimes seen in general relativity need not occur
. Furthermore, we also consider the existence conditions for some other simple anisotropic Bianchi I vacuum solutions in which the expansion in each direction is of exponential, rather than power-law behaviour and their relevance for cosmic ``no-hair'' theorems.} 
%\pacs{PACS numbers:98.80.Jk, 04.50.Kd, 04.20.Jb}

\section{\protect\bigskip Introduction}
%General introduction on higher-order theories, their motivation.
The literature contains numerous studies on generalisations of the familiar Einstein-Hilbert action of general relativity (GR) to more complicated functions of the curvature and higher-order invariants \cite{ottewill, schmidt1, schmidt2}. Motivation for these studies comes from several sources, including astronomical phenomena which are currently inadequately explained by the standard model of general relativity, such as providing a natural source of inflation in the early universe \cite{starob1}, or dark energy and the late-time acceleration of the universe's expansion \cite{Carroll:2004de, Nojiri:2006be, Nojiri:2006su, Nojiri:2006jy, Nojiri:2006gh, Amendola:2006kh}, and also attempts to include quantum behaviour in the gravitational theory \cite{stelle}. A review of one of the most common extensions to general relativity, the so-called $f(R)$ models, in which the Lagrangian is allowed to be a general function of the scalar curvature, may be found in \cite{sot}. 

%Singularities
It is of particular interest to discover the behaviour of these higher-order theories at high curvatures and it is in this limit when we might expect the influence of quantum corrections to become important. Therefore, where initial singularities are expected to involve infinities in one or more of the curvature invariants of the space-time, we expect that the addition
of higher-order terms to the Lagrangian might produce a new dominant behaviour to such singularities. The (past) stability properties of special initial isotropic cosmological singularities were investigated in \cite{midd2} for higher-order theories where the dominant term in the Lagrangian took the form $(R_{ab}R^{ab})^n$.

%Anisotropy
It is well known that contributions to the Lagrangian from terms dependent on the scalar curvature only are conformally equivalent to the presence of a minimally coupled scalar field in general relativity \cite{conformal}, however, in theories in which the gravitational Lagrangian contains higher-order curvature invariants, a much richer diversity of anisotropic behaviour is possible. For example, specific counterexamples were found in \cite{hervik} which demonstrate that the ``cosmic no-hair theorem'' of general relativity may be violated in higher-order theories. Furthermore, anisotropies diverge faster than isotropies at high curvatures and will tend to dominate the cosmological behaviour at early times. Thus, whilst the majority of previous studies of these modified theories of gravity have focussed on the behaviour of isotropic cosmologies, it is the role of anisotropy on approach to the initial singularity which we wish to investigate here.

%Kasner, BKL, Mixmaster solutions
Some of the most important anisotropic cosmological solutions in general relativity are the vacuum Kasner solutions of Bianchi type I \cite{kasner}. Since they are characterised by just a single free constant, they are geometrically special, but nevertheless they provide us with a very useful insight into the dynamics of anisotropies, since they give a good description of the evolution of more general anisotropic cosmological models over finite time intervals. The chaotic oscillatory behaviour of the spacetime on approach to the initial singularity exhibited by the Bianchi type VIII and type IX (``Mixmaster'') cosmologies can be approximated by a sequence of different Kasner epochs \cite{bkl,Misner,PhysRevLett.46.963,Barrow:1981sx,Chernoff:1983zz,Rendall:1997dc}. Provided that at least one of the Kasner exponents is negative, inhomogeneities and perturbations from the Bianchi I anisotropies will grow as the singularity is approached and force the solution to switch from one set of Kasner exponents to another. If the solution must always have at least one negative Kasner exponent, as is the situation in general relativity, then these oscillations will continue infinitely as the singularity is approached. However, in some higher-order theories of gravity \cite{deruelle, clifton1, clifton2} it may be possible for all of the Kasner indices to take positive values, whence after a sufficient (finite) number or permutations, the indices will reach such a configuration and the oscillatory behaviour will cease. Since all spatial directions will be contracting, the initial singularity will be reached monotonically.

%Previous results in this area
Previously, Kasner-type cosmological models in quadratic gravity and in Lovelock theories of gravity in higher dimensions were investigated by Deruelle \cite{deruelle}. Clifton and Barrow \cite{clifton1, clifton2} discovered the conditions for the existence of Kasner-like solutions and the exact forms of these solutions for the particular cases where the Lagrangian is an arbitrary power of one of the curvature invariants, $R, R_{ab}R^{ab}$ or $R_{abcd}R^{abcd}$. One might expect that the dynamics and the asymptotic behaviour of any solution in a more general higher-order theory would be controlled by the highest powers of the curvature in the past, and the lowest powers of the curvature in the future. However, this assumption is not necessarily accurate \cite{clifton3} and we wish to extend the investigation of \cite{clifton1, clifton2} to include more general Lagrangians. 

%What we will investigate
Thus, in this work, we wish to investigate, within this class of higher-order theories of gravity, the constraints on the Lagrangian for the existence of some simple anisotropic but homogeneous solutions of Bianchi Type I and to find all such solutions. In particular, our main focus will be to consider the possibility of anisotropic Kasner-like solutions in vacuum, and in the presence of a cosmological constant. We will also consider the properties of these solutions with respect to their relation to the behaviour of the more general Bianchi type VIII and IX cosmologies. In addition, we will also discover all solutions in these higher-order theories which are expanding anisotropically but exponentially in each of the three spatial directions.

\section{Field Equations}

In this paper we will consider theories of gravitation in which the field equations are derived from an arbitrary analytic function of the three possible linear and quadratic contractions of the Riemann curvature tensor; $R, R_{ab}R^{ab}$ and $R_{abcd}R^{abcd}$. The relevant action
is given by
\begin{equation*}
S=\int d^{4}x\sqrt{-g}\left[ \frac{1}{\chi }f(X,Y,Z)+L_{m}%
\right] ,
\end{equation*}
where $f(X,Y,Z)$ is an arbitrary function of $X, Y$ and $Z$ which are defined
$X \equiv R, Y \equiv R^{ab}R_{ab}$ and $Z \equiv R^{abcd}R_{abcd}$. Spacetime indices run from $0$ to $3$ and are denoted by roman letters, whilst Greek letters are used to denote purely spatial indices.

Unlike the situation in general relativity, which may be recovered by choosing $f=R$, the Palatini and metric formalisms are not equivalent for a general choice of $f$. In what follows, we shall restrict attention to the metric formalism. The field equations obtained by
varying the action with respect to the metric are \cite{clifton2}:
\begin{equation}
        P_{b}^{a}= \frac{\chi }{2}T_{b}^{a}\, ,
\end{equation}
where
\begin{eqnarray*}
P^{ab} & = &-\frac{1}{2}f g^{ab}+f_X R^{ab}+2f_{Y} R^{c(a}R^{b)} \, _{c}
+2f_{Z} R^{edc(a}R^{b)}\,_{cde}+(g^{ab}g^{cd}-g^{ac}g^{bd})f_{X;cd} \\
&& +\Box (f_{Y}R^{ab}) +g^{ab}(f_{Y} R^{cd})_{;cd} -2(f_{Y}R^{c(a})_;
\,^{b)}_{c} - 4(f_{Z}R^{d(ab)c})_{;cd} \: ,
\end{eqnarray*}%
where $f_X \equiv \frac{\partial f}{\partial X}$, $f_Y \equiv \frac{\partial
f}{\partial Y}$ and $f_Z \equiv \frac{\partial f}{\partial Z}$. Whilst the focus of this work is on solutions in vacuum, with $T^a_b=0$, the study can be extended in a simple way to allow the possibility of a non-zero cosmological constant, $\Lambda$, by including any such term in the gravitational part of the Lagrangian, $f(X,Y,Z)$. Furthermore, we will also consider the effects of including a perfect fluid for $f(R)$ theories of gravity.

\section{The Kasner Model}
In this paper, we consider homogeneous but anisotropic Bianchi I models, and in particular our main focus will be on those spacetimes described by the line element
\begin{equation}
ds^{2}=-dt^{2}+t^{2p_{1}}dx^{2}+t^{2p_{2}}dy^{2}+t^{2p_{3}}dz^{2} ,  \label{kas}
\end{equation} where the Kasner exponents $p_{\alpha}$ are constants and assumed to be real in order for the metric to be of physical significance. We define the useful quantities \begin{eqnarray*}
H & \equiv & p_{1}+p_{2}+p_{3}  \: ,\\
J & \equiv & p_{1}\!^2 + p_{2}\!^2 + p_{3}\!^2  \quad \text{and} \\
K & \equiv & p_{1}\!^3 + p_{2}\!^3 + p_{3}\!^3
\end{eqnarray*}

The line element (\ref{kas}) also describes the limiting case in which all three Kasner exponents are equal, $p_{1}=p_{2}=p_{3}$, and corresponds to isotropic, spatially flat Friedmann-Robertson-Walker solutions for which the scalefactor is a power-law in time. These solutions will be included for completeness. It is useful to note that for real-valued choices of $p_{\alpha}$, $J\geq \frac{H^2}{3}$, with equality if and only if the solution is isotropic.  The three relevant curvature scalars with which we will be working take the values
\begin{eqnarray}
X \equiv R & = & \frac{J-2H+H^2}{t^2} \: , \label{eq:x} \\
Y \equiv R_{ab}R^{ab} &=& \frac{J(H-1)^2+(J-H)^2}{t^4} \: , \label{eq:y}\\
Z \equiv R_{abcd}R^{abcd} &=& \frac{(3J-H^2)^2 +12J+8K(H-3)}{3t^4}  \label{eq:z} \: .
\end{eqnarray}

In general relativity, the Kasner exponents for the vacuum solution must satisfy $H=J=1$ \cite{kasner}. One solution of this is the Milne model \cite{Milne}, which without loss of generality can be described using the choice of axes such that $p_{1}=1$, $p_{2}=p_{3}=0$. However, this solution is related to Minkowski space by a coordinate transformation; if we introduce new coordinates $\tau =  t \cosh{x}, \chi=t \sinh{x}$, then the usual form of the flat Minkowski metric is explicitly recovered \cite{clifton2}. The Riemann tensor $R_{abcd}$ vanishes for Minkowski space, and therefore this is a vacuum solution in any higher-order theory of gravity of the form $f(X,Y,Z)$ for which $f(0,0,0)=0$.

For all other general relativistic Kasner solutions, one Kasner exponent must be negative, whilst the other two are positive. Thus, although the spacetime volume is expanding to the future, one of the spatial directions is contracting. Moreover, as the initial singularity is approached, inhomogeneities and deviations from the Bianchi type I anisotropies in the more general Bianchi type IX solution will grow and cause the Kasner exponents to be permuted to different values, leading to an infinite series of chaotic BKL oscillations between different Kasner epochs \cite{bkl,Misner,PhysRevLett.46.963,Barrow:1981sx,Chernoff:1983zz,Rendall:1997dc}. 

However, this is not true in general for the higher-order theories considered in this paper. In \cite{clifton1, clifton2}, Clifton and Barrow found some exact solutions, a subset of which permits all the Kasner exponents to take positive values. We will see that this such solutions exist in a much wider class of higher-order theories. In this scenario, it is possible that after a finite number of transitions between different Kasner epochs, the solution will reach a state in which all of the Kasner exponents are positive. Once this occurs, the perturbations from the Bianchi I model would not grow and the chaotic oscillations will cease. Thus, the solution will remain in this epoch and the initial singularity is then approached monotonically.

By considering those classes of theories in which only one curvature scalar - $\Phi$, say - contributes a time scale to the Lagrangian, one can solve for the time coordinate $t$ in terms of the scalar $\Phi$. In this way, time derivatives may be eliminated and the field equations may be re-written in terms of $\Phi$, the Lagrangian $f(\Phi)$, and its derivatives with respect to $\Phi$. The resultant differential equation(s) can then be solved to find all possible forms of $f$. This technique, which was also used by Dunsby et al. in $f(R)$ theories \cite{dunsby}, allows us to find all possible exact Kasner-like vacuum solutions within this general class of Lagrangians, with one exception, which may be dealt with separately. This exception is when quantity which usually determines the time scale becomes independent of time for some special choice of the parameters $p_{\alpha}$. For example, in general for the metric (\ref{kas}), the scalar curvature, $R$, is proportional to $t^{-2}$, and so in the context of $f(R)$ models one can substitute for the time coordinate $t$ using $R$, but it is necessary to consider separately those possible solutions with $R=0$.

As a consequence, we will find that the subset of metrics for which the curvature scalars are independent of time plays an important role and it is useful to discuss those metrics briefly now. Note from the expressions (\ref{eq:x}), (\ref{eq:y} and (\ref{eq:z}), the scalars $R, R_{ab}R^{ab}$ and $R_{abcd}R^{abcd}$ are only constants if and when they are zero. A special case within this class of metrics is flat Minkowski space, which may be respresented by the metric (\ref{kas}) with $p_{\alpha}=0$, that is to say $g_{ab}=\eta_{ab}$, for which the Riemann tensor, $R_{abcd}$, is identically zero, and thus $X=Y=Z=0$. This is a solution in all theories for which $f(0,0,0)=0$ and the discussion in subsequent sections will be concentrated on metrics describing curved spacetimes.

%Z=0
By expressing the Kretschmann scalar, $Z \equiv R_{abcd}R^{abcd}$, as \begin{equation*}
Z=\frac{4}{t^4}\left(p_{1}\!^{2}(p_{1}-1)^{2}+p_{2}\!^{2}(p_{2}-1)^{2}+p_{3}\!^{2}(p_{3}-1)^{2}+p_{1}\!^2 p_{2}\!^2+p_{1}\!^2 p_{3}\!^2+p_{2}\!^2 p_{3}\!^2 \right) \:,
\end{equation*}it can be seen that there are two possible real solutions to $Z=0$, given by Minkowski space, $p_{\alpha}=0$, and $p_{1}=1, p_{2}=p_{3}=0$ (plus permutations). However, as we have seen, the latter solution is the Milne model, and is related to the former by the coordinate transformation $\tau =  t \cosh{x}, \chi=t \sinh{x}$.
%Y=0
In order to satisfy $Y \equiv R_{abcd}R^{abcd}=0$, real solutions must have either $H=J=0$ or $H=J=1$. The only possibility in the former case is Minkowski space. The latter are the well-known Kasner solutions of general relativity \cite{kasner}. Thus, $Z=0$ implies that $Y=0$.

%X=0

Solutions for which the scalar curvature, $X \equiv R$, is zero require $J=2H-H^{2}$ and thus $Y=0$ implies that $X=0$. From this equation for $J$, one can see that real solutions must satisfy the constraint $J \leq 1$. Furthermore, from our definitions, $J\geq \frac{H^2}{3}$ for all real-valued choices of the constants $p_{\alpha}$, and so it is also necessary that $0\leq H \leq 3/2$. According to these constraints, we see that the spacetime volume of any solution of this form must expand no faster than $t^{3/2}$, and that the expansion rate in any one direction can be no faster than $t$. Finally, we note that solutions with $X=0$ and $H \leq 1$ cannot have all three Kasner exponents positive and, except for Minkowski space and the Milne model, at least one exponent must be negative. This is the situation in general relativity and, except for the two cases above, implies that although the spacetime volume is expanding, space must be contracting in one direction. In contrast, if $X=0$ but $1 < H <3/2$, then it is possible for all the Kasner exponents to be distinct and positive. Thus, as we have discussed, these solutions can avoid the infinite series of chaotic oscillations between different Kasner epochs on approach to the initial cosmological singularity seen in the BKL picture. The only solution with $X=0$ and $H=3/2$ is the isotropic solution with $p_{1}=p_{2}=p_{3}=1/2$.

In what follows, we shall consider in turn several commonly-studied classes of higher-order theories of gravity, obtaining the forms of the Lagrangian within these classes for which solutions of the form (\ref{kas}) exist, and the conditions which the Kasner exponents are subject to in each case.

\subsection{$f=f(R)$}
If the Lagrangian depends on the scalar curvature only, $f=f(R)$, then it is well-known to be conformally equivalent to general relativity with a minimally coupled scalar field \cite{conformal}. Thus, one does not expect a large range of anisotropic behaviour to be possible in such models; indeed for the case of quadratic corrections to the Einstein-Hilbert Lagrangian it is precisely the presence of the Ricci term, $R_{ab}R^{ab}$, which permits the existence of the anisotropic Bianchi I solutions found by Barrow and Hervik \cite{hervik}.

The vacuum field equations are $P^a_b=0$, where, for the Kasner-like metric given by (\ref{kas}), we have
\begin{eqnarray}
P^0_0 &=& -\frac{t^2 f+2(H - J)f_{R} + 2Ht \dot{f_{R}}}{2t^2} \: ,\\
P^{\alpha}_{\alpha} &=& \frac{-3t^2 f+2 H(H-1)f_R -4Ht\dot{f_R}-6t^2 \ddot{f_R}}{2t^2} \:  ,\\
P^{\mu}_{\mu}-P^{\nu}_{\nu} &=& \frac{1}{t^2}(p_{\mu} -
p_{\nu}) \left( (H-1)f_{R} + t\dot{f_R} \right )  \qquad \mbox{and} \label{eq:anisor}\\
%P^{\alpha}_{\alpha} &=& -\frac{t^2 f +2\beta_{\alpha}(1-H)f_{R}
%+2(H-\beta_{\alpha})t \dot{f_{R}}+2t^2\ddot{f_{R}}}{2t^2} \\
R &=& \frac{J-2H+H^2}{t^2} \: ,    \label{eq:r}
\end{eqnarray}where $\mu, \nu$ in equation (\ref{eq:anisor}) for the anistropic stress are not indices to be summed over, but instead are used to label the Kasner exponents, taking values from $1$ to $3$. Otherwise summation convention is used as normal. Recall that we may allow $f$ to contain a cosmological constant term, $f_{R}$ is used to denote $\frac{df}{dR}$ and overdots represent derivatives with respect to the coordinate time, $t$.

Given the form of equation (\ref{eq:anisor}), in what follows we shall consider separately the situations where the metric is isotropic and where it is anisotropic, both for vacuum and with the inclusion of a perfect fluid.

\subsubsection{Isotropic power-law vacuum solutions}
If all the Kasner exponents are equal, the solution is isotropic, with $p_{1}=p_{2}=p_{3}=\frac{H}{3}$,
$R=\frac{2H(2H-3)}{3t^2}$ and there is only one independent field equation:
\begin{equation} 3t^2 f-2H(H-3) f_{R}+ 6Ht\dot{f_{R}} = 0 \: .
\end{equation}
For a particular theory, that is to say a particular choice of $f(R)$, this equation
might appear to provide a constraint on $H$, the one free constant remaining.
However, in general, it is algebraic in $H$ \textit{and} $t$ and therefore one
will not always be able to find \textit{constant} solutions of this equation
for $H$; in fact such solutions are rare.

For models with Lagrangians of the form $f=f(R)$, it is possible to summarise the full set of possible isotropic power-law vacuum solutions, and the conditions on the function $f$ for these to be valid, using the classification in table \ref{tab:isoRv}.

\begin{table}[ht]\small 
\begin{center}
    \begin{tabular}{ l || l | p{9.5cm} }   
Class & Solution & Validity \\[2pt] \hline \hline 
\textbf{0} & $H=0$ & Minkowski space is a vacuum solution in any model with $f(0)=0$. \\[4pt]
\textbf{I} & $H=3/2$ & This solution, analagous to the radiation-dominated Friedmann universe of General Relativity, is a solution of the \textit{vacuum} field equations for any model satisfying $f(0)=0$ and $f_{R}(0)=0$. \\[4pt]
\textbf{II} & $H=\frac{3(2n-1)(n-1)}{2-n}$ & This class of solutions is possible if and only if the Lagrangian takes the form
$f=\alpha_{n} R^{n}+\alpha_{m} R^{m}$, where $n \notin \{0,5/4,2 \}$, $m=\frac{4n-5}{2(n-2)}$ and $\alpha_{n}, \alpha_{m}$ are constants, with $\alpha_n \neq 0$. These solutions are expanding to the future if $n< 1/2$ or $1<n<2$. \\
\end{tabular}\caption{Isotropic power-law vacuum solutions in $f(R)$ gravity.}\label{tab:isoRv}
\end{center}
\end{table}

The class II solutions were found previously in \cite{clifton1} in the context of models where the Lagrangian is a power of the scalar curvature, $f=R^n$. However, it is important to note that for all such models with $n>1$, the solution of class I, which has zero scalar curvature, also solves the vacuum field equations, a fact which was not stressed in \cite{clifton1}. Furthermore, for any model in which the Lagrangian may be written as a power series in $R$, a non-zero linear term (i.e. an Einstein-Hilbert term) in the series precludes the possibility of isotropic power-law vacuum solutions other than Minkowski space.

\subsubsection{Isotropic power-law solutions with a perfect fluid}
Let us now consider a universe filled with a comoving perfect fluid, with equation of state $p= w \rho$, and energy density evolving as $\rho (t) = \rho_0 t^{-H(1+w)}$, where $\rho_0$ is a constant. The possible isotropic power-law solutions of the field equations for a Lagrangian of the form $f=f(R)$ are similar to those found in the vacuum case and are summarised in table \ref{tab:isoRf}.

\begin{table}[ht]\small 
\begin{center}
    \begin{tabular}{ l || l | l | p{6cm} } 
Class & Solution & Equation of state & Validity \\[2pt] \hline \hline 
\textbf{I} & $H=3/2$ & $w=1/3$ & This is a solution in any model satisfying $f(0)=0$ and $f_{R}(0)= constant \neq 0$.  \\[4pt]
\textbf{II} & $H=\frac{2n}{1+w}$ & $w \neq -1$  & This class of solution exists iff the Lagrangian is a power of the scalar curvature, $f=\alpha_{n} R^{n}$, where $n \neq 0$ and $\alpha_{n}$ is a constant. \\[4pt]
\textbf{III} & $H=\frac{3(2n-1)(n-1)}{2-n}$ & $w=-1+\frac{2p(2-n)}{3(2n-1)(n-1)}$ & This is a solution iff the Lagrangian takes the form
$f=\alpha_{n} R^{n}+\alpha_{m} R^{m}+\alpha_{p}R^{p}$, where $n \notin \{0,5/4,2 \}$, $m=\frac{4n-5}{2(n-2)}$ and $\alpha_{p}\neq0 , \alpha_{n}$ and $\alpha_{m}$ are constants. For power-law Lagrangians ($\alpha_m=\alpha_n=0$), this solution reduces to a subset of the class II solutions. \\
\end{tabular}\caption{Isotropic power-law solutions in $f(R)$ gravity with a comoving perfect fluid.} \label{tab:isoRf}
\end{center}
\end{table}

The solution of class I is analagous to the radiation-dominated Friedmann universe of General Relativity, and also to the class I \textit{vacuum} solution when an extra Einstein-Hilbert term is added to the Lagrangian.

It is apparent that the vacuum solutions found in the previous section correspond to fluid-filled solutions with an appropriate extra curvature term in the Lagrangian.

\subsubsection{Anisotropic vacuum solutions}
We have seen that if the Lagrangian is a function of the scalar curvature only, $f=f(R)$, there is a strong constraint on its form for the
existence of exact isotropic power-law solutions. We shall now turn our attention to the situation
for anisotropic Kasner-like solutions within these models, the main focus of this paper. In this case, the vacuum field equations can be reduced to \begin{eqnarray}
(H-1)f_{R}+t\dot{f_{R}} &=& 0 \label{c1} \: , \\
f &=& 0 \label{c2} \: ,\\
Rf_{R} &=& 0  \label{c3} \: .
\end{eqnarray}

If the arbitrary function $f$ is non-trivial and algebraic in the scalar curvature, $R$, then the only possible solutions for $R$ of $f(R)=0$ must be constants, and therefore any such solution satisfies $R=0$ and thus $J=2H-H^2$. Recall that from our definitions, solutions with zero scalar curvature for which the constants $p_{\alpha}$ take real values must satisfy the constraints $0\leq H \leq 3/2$ and $J \leq 1$. According to these constraints, we see that the spacetime volume of any solution of this form must expand no faster than $t^{3/2}$, and that the expansion rate in any one direction can be no faster than $t$.

Whilst $f(0)=0$ is a necessary condition for a $f(R)$ model to contain anisotropic Kasner-like solutions in vacuum, and this also implies that equation (\ref{c3}) is satisfied at $R=0$, it is not sufficient, since $f_R(0)$ is model-dependent. Thus it is not guaranteed that there will be solutions of equation (\ref{c1}) with \textit{constant} $H$ in models where $f_R(0)$ diverges; in this case, solutions will only exist if the divergent part of the Lagrangian is a power of the scalar curvature. If, however, $f_{R}(0)$ is zero, then equation (\ref{c1}) is satisfied trivially. If $f_R(0)$ is a non-zero constant, as is the case in general relativity, defined by $f(R)=R$, then this gives the extra constraint $H=1$, whence $J=1$ and the only possible solutions are those of general relativity.

To exemplify the situation, we could consider a Lagrangian which can be expanded as a power series about $R=0$. A non-zero linear term - an effective Einstein-Hilbert term - in the series precludes the possibility of anisotropic Kasner-like vacuum solutions other than the general relativistic one, with $H=J=1$, whilst a non-zero constant term - an effective cosmological constant - would preclude the possibility of these solutions altogether.

A summary of the conditions that must be satisfied by the model in order for solutions of this kind to exist and classification of the possible solutions is given in table \ref{tab:kasR}.

\begin{table}[ht]\small 
\begin{center}
    \begin{tabular}{ l || p{3cm} | p{9cm} } 
Class & Solution constraints & Validity \\[2pt] \hline \hline 
\textbf{I} & $J=2H-H^2$ & These solutions are possible if $f(0)=0$ and $f_{R}(0)=0$. The exponents $p_{\alpha}$ are real provided $0\leq H \leq 3/2$. \\[4pt]
\textbf{II} & $H=J=1$  & If the model satisfies $f(0)=0$, but $f_{R}(0)$ is a non-zero constant, then the extra constraint from equation (\ref{c1}) means that only this subset of the first class of solutions is possible. These are the solutions for the case of general relativity, $f(R)=R$. \\[4pt]
\textbf{III} & $H=2n-1$, \newline $J=(2n-1)(3-2n)$ & This subset of the first class of solutions is relevant if $f(0)=0$, but $f_{R}(0)$ diverges due to a term which is a power of the scalar curvature. Thus, the Lagrangian is required to be of the form $f=\alpha R^n+ \hat{f}(R)$, with $0<n<1$ and $\hat{f}(0)=\hat{f}_R(0)=0$, although $\hat{f}$ need not be identically zero. These solutions are complex and of limited physical interest if $n < 1/2$. \\
\end{tabular} \caption{Kasner-like vacuum solutions in $f(R)$ gravity.} \label{tab:kasR}
\end{center}
\end{table}

The first class of solutions contains a subset for which all the Kasner indices may be positive; this is possible if $1<H<3/2$.  We have seen that the existence of such solutions means that the infinite series of chaotic oscillations on approach to the initial singularity in the more general Bianchi type IX Mixmaster universe can be avoided. 

The third class of these solutions are relevant for models in which the Lagrangian is an arbitrary power of the scalar curvature, $f=\alpha R^n$, and were found in that context by Clifton and Barrow \cite{clifton1}. However, for $n>1$, we have found here that there are additional exact Kasner-like solutions corresponding to $R=0$ which do not necessitate that $H=2n-1$ and are subject only to $J=2H-H^2$.

\subsubsection{Anisotropic solutions with perfect fluid}
It is interesting to also include the possibility of a comoving perfect fluid with a barotropic equation of state, $p=w \rho$. The energy density of the fluid is given by
\begin{equation}
 \rho = \rho_0 t^{-H(1+w)} \: .
\end{equation}
If the fluid is comoving, the spatial part of the energy-momentum tensor is isotropic and so equation (\ref{c1}) still holds. This can be solved to give $f_R \propto t^{1-H}$ and the remaining field equations simplify to
\begin{eqnarray}
 \rho(1+w) &=& -R f_R  \: ,\\
w \rho &=& -f/2 \:  .
\end{eqnarray}
The only solutions of these equations with constant non-zero energy density correspond to a cosmological constant, which was considered together with the vacuum case in the previous section. For all other fluid-filled solutions, $\rho \propto t^{-H(1+w)}$ and $R \propto t^{-2}$, and thus the field equations can only have a solution of this sort if the Lagrangian is a power of the scalar curvature, $f(R)=R^n$, for some $n \in \mathbb{R} \backslash \{0\}$, and $R \neq 0$. For $f=R^{1/2}$, the right hand sides of the above equations are identically equal, hence to allow a non-zero energy density, we further require $n \neq 1/2$.

For all $n \neq 1/2$, these equations have the solution $H=2n-1, w=(2n-1)^{-1}$. Since these solutions need $R \neq 0$, it is also required that $J \neq (2n-1)(3-2n)$, but otherwise it is unrestricted. 

\subsubsection{Examples of solutions for specific choices of $f(R)$}
To summarise the results of the previous sections, in table \ref{tab:existR} we consider some of the more commonly-studied choices for the Lagrangian, $f(R)$, and state whether these models contain Minkowski space, isotropic power-law solutions, and exact anisotropic Kasner-like solutions, both in vacuum and in the presence of a comoving perfect fluid, according to the classifications given in the preceeding sections. We include the choices of power-law Lagrangians \cite{clifton1}, and two exact $f(R)$ models which have recently been proposed, by Starobinsky \cite{starob2}, and by Hu and Sawicky \cite{hu}. These models are of particular interest since they provide viable cosmologies and evade the known constraints on the form of $f(R)$ from solar system tests. They are given by
\begin{eqnarray}
f(R) = f_{\text{Star}}(R) & \equiv & R-\lambda R_{0}\left(1-\left(\frac{1}{1+(R/R_{0})^2}\right)^{n}\right) \quad \text{and}\\
f(R) = f_{\text{Hu}}(R) & \equiv & R+\lambda R_{0}\frac{(R/R_{0})^n}{1+\alpha (R/R_{0})^n}
\end{eqnarray}
respectively. The parameter $n$ is taken to be positive for both the Starobinsky and the Hu-Sawicky models, in order to ensure their viability with observations.

\begin{table}[ht]\small 
\begin{center}
\begin{tabular}[ht]{l||c|c|c|c|c}
 $f(R)$ & Minkowski & \multicolumn{2}{c|}{Power-law} & \multicolumn{2}{c}{Kasner}
\\
 & & in vacuum & with fluid & in vacuum & with fluid \\ \hline\hline
 $R$ & $\checkmark$ & $\times$ & Class I, II & Class II & $\checkmark$  \\
 $R+\Lambda$ & $\times$ & $\times$  & $\times$    & $\times$ & $\times$ \\
 $R+\alpha R^{2}$ & $\checkmark$ & $\times$ & Class I,III  & Class II & $\times$ \\
 $R+\alpha /R$ & $\times$ & $\times$ & Class III  & $\times$ & $\times$ \\
 $R^{n}, n<0$ & $\times$ & $\times$ & Class II & $\times$ & $\checkmark$  \\
 $R^{n}, 0<n<\frac{1}{2}$ & $\checkmark$ & Class II & Class II & $\times$ & $\checkmark$ \\
 $R^{1/2}$ & $\checkmark$ & $\times$ & $\times$ & $\times$ & $\times$ \\
 $R^{n}, \frac{1}{2}<n<1$ & $\checkmark$ & Class II & Class II & Class III & $\checkmark$  \\
 $R^{n}, n>1$ & $\checkmark$ & Class I, II & Class II & Class I & $\checkmark$  \\
   $\exp{(R/R_{0})}$ & $\times$ & $\times$ & $\times$     & $\times$ & $\times$ \\
 $f_{\text{Star}}$ & $\checkmark$ & $\times$ & Class I     & Class II & $\times$ \\
 $f_{\text{Hu}}, n \geq 1$ & $\checkmark$ & $\times$ & Class I     & Class II & $\times$ \\
   $f_{\text{Hu}}, 1>n>0$ & $\checkmark$ & $\times$ & $\times$     & $\times$ & $\times$ \\
   $f_{\text{Hu}}, n<0$ & $\times$ & $\times$ & $\times$     & $\times$ & $\times$    
\end{tabular} \caption{The existence of power-law and Kasner-like solutions in various models of $f(R)$ gravity.}   \label{tab:existR}
 \end{center}
\end{table}

\subsubsection{Remarks}
Friedmann-Robertson-Walker power-law solutions in $f(R)$ gravity with a perfect fluid were investigated in \cite{dunsby}. There, it was claimed that the only possible form of Lagrangian which admits these solutions and has the correct general relativistic limit is a power-law, $f=R^n$. In fact, there is another class of isotropic power-law solution (type I in our classification) corresponding to a radiation-filled universe, which is possible in any theory for which $f(0)=0$ and $f_{R}(0)$ is constant, however comoving perfect fluids with other equations of state are not possible. Furthermore, we have seen here that it is only the $R^{n}$ Lagrangians which allow anisotropic Kasner-like solutions with a perfect fluid. We can see, therefore, that these $R^{n}$ theories of gravity are special in admitting this sort of solution. Some Bianchi type I, III and Kantowski-Sachs solutions in $f(R)$ gravity have also been investigated recently by Farasat Shamir \cite{Shamir:2010ee}.

Anisotropic singularities of Bianchi Type I were recently studied in the context of more general Lagrangians of the type $f(R, \phi, \chi)$, with $\chi = -\frac{1}{2} g^{ab} \partial _{a} \phi \partial _{b} \phi$ \cite{saa}, with particular focus on the anistropic instabilities related to the existence of the hypersurface $\frac{\partial f}{\partial R} =0$ and solutions being able to cross this surface, leading to questions about the viability of these models. Here, it has been shown that exact anisotropic Kasner-like solutions in $f(R)$ theories must have $R=0$ and although they can live on the hypersurface defined by $\frac{\partial f}{\partial R} =0$, they cannot cross it.

\subsection{$f=f(R^{ab}R_{ab})$}
Let us now consider the case where the Lagrangian is a function of the Ricci invariant only. For isotropic cosmologies, the contributions to the field equations from the simplest such term, $R_{ab}R^{ab}$, are proportional to those from a term quadratic in the scalar curvature, $R^2$. However, this is not true more generally, and the Ricci term allows much more diverse anisotropic behaviour \cite{hervik}.

  For this class of theories, the relevant field equations for the metric (\ref{kas}) in vacuum are $P^{a}_{b} = 0 $, where
\begin{eqnarray}
P^0_0 & = & \frac{1}{2t^4}\left(-t^4 f +
 2 (-H^2 + H^3 - HJ +2J^2-J) f_{Y} + 2(H^2 + J - 2 HJ) t \dot{f_{Y}} \right)\: ,\label{eqY1}
\\
P^{\mu}_{\mu}-P^{\nu}_{\nu} &=& \frac{1}{t^4}(p_{\mu} -
p_{\nu}) \left(2 (H-3)(J - 1) f_{Y} + (-4 + 2 J + 3 H - H^2)t
\dot{f_Y}+(1-H) t^2 \ddot{f_{Y}} \right)  \label{eqY3} \: , \end{eqnarray}
where as before $\mu, \nu$ in equation (\ref{eqY3}) for the anistropic stress are not indices to be summed over, but are used here as labels, taking values from $1$ to $3$. Otherwise summation convention is used as normal. Recall that $f_{Y}$ is used to denote $\frac{df}{dY}$ and overdots represent derivatives with respect to the time coordinate $t$. The Ricci term is given by\begin{equation}
Y \equiv R_{ab}R^{ab} = \frac{J(H-1)^2+(J-H)^2}{t^4}  \, .\label{Y}
\end{equation}

It is useful to recall our earlier observation that for real-valued choices of the constants $p_{\alpha}$, the Ricci term, $Y$, takes non-negative values, and is zero if and only if $H=J=1$ or $H=J=0$.

\subsubsection{Isotropic power-law solutions}
For $f=f(Y)$, and the special case of an isotropic metric, $p_{1}=p_{2}=p_{3}=H/3$, so that $J=H^2 /3$, the
equations reduce to:
\begin{equation}
-9 t^4 f + 4H^2((-6 + 3 H + H^2) f_Y - 3 (H-2) t \dot{f_Y}) = 0 \:  .
\end{equation}
By an argument analagous to that used before for the $f(R)$ theories, $f(Y)$ can be zero for all times only if $Y$ is a constant, and therefore zero. Using (\ref{Y}), it is clear that Minkowski space is the only real isotropic power-law solution for which $Y=0$. If $Y$ is non-zero, then one can eliminate the time variable $t$ and instead consider this equation as a differential equation in $Y$. It may then be integrated to find that, in order to admit solutions of this sort, the Lagrangian is required to be of the form of a power of the Ricci invariant, or possibly a sum of two such terms.

We can now summarise the existence conditions for vacuum solutions of this type for theories with Lagrangians of the form $f=f(Y)$, using the definitions $H_{\pm}(m)\equiv \frac{3- 9 m + 12 m^2 \pm \sqrt{3(-1 + 10 m - 5 m^2 - 40 m^3 + 48m^4)}}{2(1-m)}$ and $n_{\pm}(m) \equiv
\frac{2m-1}{2(m-1)}+\frac{1}{2 + 10 m - 24 m^2 \mp
 2\sqrt{3(-1 + 10 m - 5 m^2 - 40 m^3 + 48 m^4)}}$. This is found in table \ref{tab:isoY}.

\begin{table}[ht]\small 
\begin{center}
    \begin{tabular}{ l || l | p{9cm} } 
Class & Solution & Validity \\[2pt] \hline \hline 
\textbf{0} & $H=0$ & Minkowski space is a solution in any model with $f(0)=0$. \\[4pt]
\textbf{I} & $H=3/2$ & This is a solution if the Lagrangian is linear in $Y$, $f(Y) = \alpha Y$, where $\alpha$ is a non-zero constant.  \\[4pt]
\textbf{IIa} & $H=H_{+}(m)$ & This solution is possible if and only if the Lagrangian takes the form
$f=\alpha_{m} Y^{m}+\alpha_{n} Y^{n}$, where $m \neq 1$, $n=n_{+}(m$ and $\alpha_{n}, \alpha_{m}$ are constants, with $\alpha_m \neq 0$. These solutions are expanding to the future if $m<1$ \\[4pt]
\textbf{IIb} & $H=H_{-}(m)$ & This solution is possible if and only if the Lagrangian takes the form
$f=\alpha_{m} Y^{m}+\alpha_{n} Y^{n}$, where $m \neq 1$, $n=n_{-}(m)$ and $\alpha_{n}, \alpha_{m}$ are constants, with $\alpha_m \neq 0$. These solutions are expanding to the future if $m<1/4$ or $m>1/2$.\\
\end{tabular} \caption{Isotropic power-law vacuum solutions in $f(R_{ab}R^{ab})$ gravity.} \label{tab:isoY}
\end{center}
\end{table}

As with the situation for the Lagrangian $f=R^2$, we see that the radiation-dominated Friedmann universe of General Relativity is a vacuum solution of the higher-order quadratic theory. 

Whilst $H_{+}$ is unbounded as $n\rightarrow \pm \infty$, $H_{-}$ is bounded and tends to $2$ in both these limits. For power-law Lagrangians in the Ricci term, both class IIa and IIb solutions are valid; these were found in \cite{clifton2} and their stability on approach to the initial singularity under small perturbations of the metric was previously studied in \cite{midd2}.

\subsubsection{Anisotropic vacuum solutions}
For anisotropic Kasner-like solutions, we again consider separately the cases $Y=0$ and $Y \neq 0$, so that in the latter scenario, we can replace the time variable $t$ and treat the field equations (\ref{eqY1}-\ref{eqY3}) as differential equations in the Ricci term, $Y \equiv R_{ab}R^{ab}$. Similarly to the situation for isotropic power-law metrics in these theories, anisotropic solutions of the field equations with $Y \neq 0$ can only exist if the Lagrangian for the theory is a power of the Ricci term, with $f=\alpha Y^n$.

The conditions for the existence of anisotropic Kasner-like solutions within this class of theories and the constraints that must be satisfied by the Kasner indices may be summarised as in table \ref{tab:kasY}.

\begin{table}[ht]\small 
\begin{center}
    \begin{tabular}{ l || p{4cm} | p{8cm} } 
Class & Solution constraints & Validity \\[2pt] \hline \hline 
\textbf{I} & $H=J$ & There is a family of anisotropic solutions of this sort if $f(Y)=\alpha Y^{1/2}$.  The Kasner exponents $p_{\alpha}$ are real provided that $0\leq H \leq 3$. \\[4pt]
\textbf{II} & $H=J=1$  & These solutions require that $f(0)=0$, and also that $f_{Y}(0)$ either converges to a constant or diverges slower than $Y^{-1/2}$. \\[4pt]
\textbf{III} & $H=(1-2n)^2$, \newline $J=(1-2n)(1-6n+4n^2)$ & These are solutions for Lagrangians of the form $f(Y) = \beta Y^n$. In order for the Kasner exponents to be real, it is needed that $(1-2n) (1- 6n + 4n^3) \geq 0$. \\
\end{tabular} \caption{Kasner-like vacuum solutions in $f(R_{ab}R^{ab})$ gravity.} \label{tab:kasY}
\end{center}
\end{table}

The first class of solutions have real Kasner exponents provided $0\leq H \leq 3$. The volume of these solutions can thus expand no faster than $t^{3}$, but the individual exponents must lie in the range $\frac{1}{2}\left( 1-\sqrt{3}\right) \leq p_{\alpha} \leq \frac{1}{2}\left( 1+\sqrt{3}\right)$, and so the expansion may be faster than the speed of light in a particular direction.

The second class of solutions are the general relativistic solutions, with $H=J=1$; we have seen that these are the only anisotropic metrics of Kasner type for which the Ricci term is zero. It is interesting to compare these with the class II anisotropic vacuum solutions found in the previous section for $f(R)$ models, which also have $H=J=1$. In that context the solutions require that $f_{R} (0)$ is constant, and thus $f \sim R$ as $R \rightarrow 0$, whilst here $f \sim Y^{1/2}$ (or higher powers) as $Y \rightarrow 0$.

The third class are the solutions found in \cite{clifton2}, in which theories where the Lagrangian is a power of the Ricci term were previously investigated. The condition that the third class of solutions have real exponents is satisfied if either $n_{1} \leq n \leq n_{2}$ or $1/2 \leq n \leq n_{3}$, where $n_{1}, n_{2}$ and $n_{3}$ are the roots of $(1- 6n + 4n^3)=0$, chosen such that $n_{1}<n_{2}<n_{3}$. These roots have approximate numerical values $n_{1}\approx -1.30,  n_{2} \approx 0.17$ and $n_{3} \approx 1.13$. These boundaries correspond to isotropic solutions and one can see their relevance for the existence of isotropic solutions in \cite{midd2}. For $n>0$, both $H$ and $J$ must be less than unity, and so for positive $n$ the expansion of the solution cannot accelerate. However, $H>1$ for these solutions if $n<0$, and in particular for models with $ n_{1} \leq n \leq \frac{1}{2}\left( 1-\sqrt{3}\right)$, $H$ will be greater than $3$ and so the volume of the spacetime must increase faster than $t^{3}$ and undergo an accelerated expansion. Whilst this third class of solutions includes the special case of quadratic gravity defined by $n=1$, it is easy to see that the conditions give the second class of solutions, $H=J=1$, and there is no additional set of solutions in this case. 

%These are the solutions found in \cite{clifton2}, in which theories where the Lagrangian is a power of the Ricci term were previously investigated. 
%\begin{itemize}
%\item We have seen that the only metrics of anisotropic Kasner type for which the Ricci term is zero are the general relativistic solutions, with $H=J=1$. These solutions require that $f(0)=0$, and also that $f'(0)$ either converges to a constant or diverges slower than $Y^{-1/2}$.
%\item If $f(Y)=\alpha Y^{1/2}$, then there is a family of anisotropic solutions specified by $H=J, 0< H \leq 3$.
%\item If $f(Y) = \beta Y^n$, then there are solutions with $H=(1-2n)^2, J=(1-2n) (1-6n+4n^2)$. %The requirement that $J\geq H^{2}/3$ implies $(1-2n) (1- 6n + 4n^3) \geq 0$, i.e. $-1.30084 \approx n_{1} \leq n \leq n_{2} \approx 0.169938$ or $1/2 \leq n \leq n_{3} \approx 1.1309$. NB These boundaries should correspond to exact isotropic solutions - recall previous paper \cite{midd2}. Whilst this includes the theory defined by $f(Y)=Y$, it is easy to see that the conditions give $H=J=1$. These are the solutions found in \cite{clifton2}, in which theories where the Lagrangian is a power of the Ricci term were previously investigated.
%\end{itemize}

\subsubsection{Examples for specific $f(R^{ab}R_{ab})$}
We are now able to summarise the possible vacuum Kasner-like solutions for various choices of Lagrangian of the form $f(Y)$. In table  \ref{tab:existRic}, we consider several more common examples of this type of model and detail whether they allow isotropic power-law or Kasner-like solutions in vacuum, according to the classification systems we have used in the preceeding sections. In this table, use is also made of the definitions given in the previous section; recall that $n_{1}, n_{2}$ and $n_{3}$ are the roots of $(1- 6n + 4n^3)=0$, chosen such that $n_{1}<n_{2}<n_{3}$, and have approximate numerical values $n_{1}\approx -1.30,  n_{2} \approx 0.17$ and $n_{3} \approx 1.13$.

\begin{table}[ht]\small 
\begin{center}
\begin{tabular}[ht]{l||c|c|c}
 %\centering
 {$f(Y)$} & Minkowski & Power-law & Anisotropic Kasner \\ \hline\hline
 $Y$ & $\checkmark$ & Class I & Class II \\
 $Y+\Lambda$ & $\times$ & $\times$      & $\times$ \\
 $Y^{n}, n<n_{1} $ & $\times$ & Class II & $\times$ \\
 $Y^{n}, n_{1}<n<0 $ & $\times$ & Class II & Class III \\
 $Y^{n}, 0<n<n_{2} $ & $\checkmark$ & Class II & Class II,III \\
 $Y^{n}, n_{2}<n<\frac{1}{2} $ & $\checkmark$ & Class II & Class II \\
 $Y^{1/2}$ & $\checkmark$ & Class II   & Class I \\
 $Y^{n}, \frac{1}{2}<n<n_{3}, n\neq 1 $ & $\checkmark$ & Class II & Class II, III \\
 $Y^{n}, n_{3}<n$ & $\checkmark$ & Class II  & Class II \\
 $Y+\alpha Y^2$ & $\checkmark$ & $\times$      & Class II \\
 $exp(Y/Y_{0})$ & $\times$ & $\times$      & $\times$ \\
 $sin(Y/Y_{0})$ & $\checkmark$ & $\times$      & Class II 
\end{tabular}\caption{The existence of power-law and Kasner-like vacuum solutions in various models of $f(R_{ab}R^{ab})$ theories of gravity.}     \label{tab:existRic}
\end{center}
\end{table}
%*other than Minkowski space.

\subsection{$f=f(Z)$}
For functions of the Kretschmann scalar, $Z \equiv R_{abcd}R^{abcd}$, only, the field equations for the metric (\ref{kas}) in vacuum reduce to $P^{a}_{b}=0$, where
\begin{eqnarray}
 P^0_0 %&=& \frac{1}{2}\left(-f+Z \left(f_Z-\frac{3t \dot{f_Z}}{H-3}\right)+\frac{H^4 + 8 H J - 6 H^2 J + 3 J (-4 + 3 J)}{t^4}\left(f_Z+\frac{t \dot{f_Z}}{H-3}\right)\right)\\
 &=& -\frac{1}{2}f +2Z f_Z+\frac{4}{t^4}(J-K)\left((H-3)f_Z+t \dot{f_Z}\right)\\
% P_i^i &=& -2f+2Z f_Z +\frac{4}{t^4}\left((H-3) (-2 H + H^2 + J) f_Z + (2 H^2 + 3 J - H (4 + J)) t \dot{f_Z} + (H - J) t^2
% \ddot{f_Z}\right) \\
% P_{\alpha}^{\alpha} &=& -\frac{3}{2}f+\frac{4}{t^4}\left((H-3) (H^2-2 H + K) f_Z + \left((H-2)(2H-J)+K \right) t \dot{f_Z} + (H - J) t^2
% \ddot{f_Z}\right) \\
P^{\mu}_{\mu}-P^{\nu}_{\nu} &=& 2 \frac{(p_{\mu}-p_{\nu})}{t^4} \biggl(2 \left(H-p_{\mu}-p_{\nu}\right) t \left((H-2) \dot{f_Z}+t \ddot{f_Z}\right) + (3-H) (4 -2H + H^2- 3 J)f_Z \nonumber \\
&& + (-8 + 8H - 3 H^2+ 3 J)t\dot{f_Z} + 2 (1 - H) t^2 \ddot{f_Z}\biggr) \: , \qquad \text{with} \label {eq:zaniso}\\
Z %&=& \frac{4 J +(3J-H^2) (J+H^2-4H)+8 a b c(H-3)}{t^4} \\
&=& \frac{(3J-H^2)^2 +12J+8K(H-3)}{3t^4} \:  .
\end{eqnarray} Once again $\mu, \nu$ in equation (\ref{eq:zaniso}) for the anisotropic stress are not indices to be summed, but are used to label the Kasner exponents, taking values from $1$ to $3$. Otherwise summation convention is used as normal. Recall that $f_{Z}$ denotes $\frac{df}{dZ}$ and overdots are used to represent derivatives with respect to the time coordinate $t$.

Recall also that, by expressing $Z$ explicitly in terms of the Kasner exponents, it can be seen that the only possible real solutions of $Z=0$ are Minkowski space and the Milne model, which may be defined by $p_{1}=1, p_{2}=p_{3}=0$ and is related to the Minkowski solution by the coordinate transformation $\tau =  t \cosh{x}, \chi=t \sinh{x}$. Thus, for all other Kasner-like metrics with $Z \neq 0$, it is possible as before to substitute for the time variable $t$ using $Z$ and integrate to find the requirements on the Lagrangian $f=f(Z)$ in order to permit such metrics as solutions of the field equations. Using this method, it is found that in order for Kasner-like solutions to exist in vacuum, the Lagrangian must be a power of the scalar $Z$, or possibly a sum of two terms if they permit the same values of the Kasner exponents $p_{\alpha}$. Such power-law Lagrangians were first investigated in \cite{clifton2}.

\subsubsection{Anisotropic solutions}
For the previous two types of models that were considered, with $f=f(R)$ or $f=f(R^{ab}R_{ab})$, there was only one independent equation for the anisotropic stress. In contrast, for models with Lagrangians that are functions of the Kretschmann scalar, $f=f(R^{abcd}R_{abcd})$, there are in general \emph{two} independent equations for the anisotropic stress. This appears to be due to the fact that the Kretschmann scalar cannot be expressed solely in terms of the parameters $H$ and $J$, unlike the scalar curvature and Ricci scalar. For the cases of interest, where the Lagrangian is specified by $f= \alpha Z^n$, with $n \neq 0$, these simplify to
\begin{eqnarray}
0 &=& (p_{1}-p_{2}) (1+H-4n)(-4-3J + H(6+H-8n)+ 8n+8p_{3} (n-1)) \: ,\\
0 &=& (p_{2}-p_{3}) (1+H-4n)(-4-3J + H(6+H-8n)+ 8n+8p_{1} (n-1)) \: .
\end{eqnarray}
Together, these imply that 
\begin{equation}
0 = (p_{1}-p_{2})(p_{1}-p_{3})(p_{2}-p_{3})(1+H-4n)(n-1) \: , \end{equation} and so fully anisotropic solutions with no two Kasner indices equal can be possible only if either $n=1$ or $H=4n-1$. However, substituting $H=4n-1$ into the remaining independent field equation $P^1_1=0$ leads to the further requirement that $Z=0$, for which we have seen that the only real solutions are Minkowski spacetime and the Milne model. The only other possibility for a fully anisotropic solution with $p_{\alpha}$ all real and distinct is in the theory with $n=1$, i.e. $f=\alpha Z$, a particular case of the commonly studied class of quadratic gravity theories, which will be discussed in more generality in the next section. Using the remaining independent field equation, such anisotropic solutions must have $H=J=1$. Recall that these were also solutions for the other special cases of quadratic gravity considered in the previous sections; $f=\alpha R^{2}$ and $f=\alpha Y$.

\subsubsection{Locally rotationally symmetric solutions}
Unlike the situation for the previous two classes of model, with $f=f(R)$ or $f=f(R^{ab}R_{ab})$, there are, in general, two equations for the anisotropic stress in those models with $f=f(R^{abcd}R_{abcd})$. Consequently, in addition to the isotropic and fully anisotropic metrics one must also consider the scenario of locally rotationally symmetric (LRS) solutions which without loss of generality are described by $p_{2}=p_{3}=(H-p_{1})/2$, $H\neq 3p_{1}$.

For these LRS Kasner-like metrics, in the context of power-law Lagrangians, $f=\alpha Z^n$, the field equations may be reduced to the two independent equations;
\begin{eqnarray}
0&=& 8 - 8 p_{1} + 9 p_{1}\!^2 - 4 H - 6 p_{1} H + H^2 + 8n(H+p_{1}-2) \label{eq:lrs1}\\
0 &=&  3 H^3 (n-1) - H^2 (4-23n-8n^2) +
     2 H (1-2n)(20 -25n + 32 n^2) \label{eq:lrs2}\\ && + (2n-1) (47-148n+128n^2) 
-     p_{1} (n-1) ((5 - 8 n) (13 - 16 n) - H (51 - 60 n)) \: . \nonumber \end{eqnarray}

For all $n$, these equations are solved by $H=p_{1}=1$, giving $p_{2}=p_{3}=0$; this is the subset of the anisotropic solutions found in the previous section with $H=J=1$ which corresponds to the Milne model. For $n=1$ there are no other LRS solutions of this form. For $n \neq 1$, the second of these equations is linear in $p_{1}$; substituting the solution one obtains for $p_{1}$ into the first equation gives a sixth order polynomial equation for $H$ in terms of $n$. However, factoring out the root $H=1$  leaves the fifth order polynomial:
\begin{eqnarray*}
 0 &=& H^5 (n-1)^2 + H^4 (1-n) (15-41n + 8 n^2) +
   H^3 (38 - 216 n + 283 n^2 + 56 n^3 - 80 n^4) \\ 
& & +
   H^2 (-346 + 2156 n - 5421 n^2 + 7032 n^3 - 4560 n^4 + 896 n^5) \\
&& + H (1-2n)^2 (653 - 2594 n + 3848 n^2 - 2688 n^3 + 1024 n^4)  \\
&& - (1-2n)^2 (361 - 2104 n + 4768 n^2 - 4992 n^3 + 2048 n^4) \:  . 
\end{eqnarray*}

Consequently, in addition to the Milne solution, there must always be at least one locally rotationally symmetric Kasner-like solution with real values for $p_{1}$ and $H$ for any real choice of $n$, other than $n=1$. For example, when $n=1/2$, there is a solution given by $p_{1}=1, H=5$, giving $p_{2}=p_{3}=2$, whilst for $n=7/8$ there are three real solutions, one of which is given by $p_{1}=-1/2, H=1/2$, so that $p_{2}=p_{3}=1/2$. Thus it is dependent upon the choice of $n$ as to whether the signs of the Kasner exponents must all be positive or not. These solutions are new and were not found in \cite{clifton2}.

\subsubsection{Isotropic power-law solutions}
In order to complete the discussion of this section, we consider the isotropic metrics, with $p_{1}=p_{2}=p_{3}=H/3$. Except for Minkowski spacetime, which is a solution in any theory for which $f(0)=0$, the relevant equation for these metrics in vacuum in the context of a power-law Lagrangian, $f=\alpha Z^n$, is
\begin{equation}
0=-1+6n-8 n^2 + \frac{2}{3}(1 - 2 n + 4 n^2) H + \frac{2}{9} (n - 1) H^2 \: . \label{eq:isoz}
\end{equation} 
For $n=1$, there is only one solution, with $H=3/2$. For $n \neq 1$, there are two solutions, given by
\begin{equation*}
H=H_{\pm} \equiv \frac{ \left(-1 + 2 n -4n^2 \pm \sqrt{-1 +10n - 16 n^2 + 16n^4}\right)}{6(n-1)} \: ,
\end{equation*} which are real if $n\leq n_{1} \equiv -\left(\sqrt{5} + \sqrt{3 + 2 \sqrt{5}}\right)/4 \approx -1.24$ or $n\geq n_{2}\equiv \left(-\sqrt{5} + \sqrt{3 + 2 \sqrt{5}}\right)/4 \approx 0.12$.

Finally, we note that if $H \neq 3$, equation (\ref{eq:isoz}) is quadratic in $n$ and so two values of $n$ correspond to the same value of $H$ and so Lagrangians which are a sum of two powers related by this equation also contain an isotropic power-law solution. Now $H=3$ for $n=\frac{1}{4}$ and thus, for $f=\alpha_{n} Z^n+\alpha_{m} Z^m$, with $\alpha_{n},\alpha_{m} \neq 0, (4n-1)(n-1)\neq 0$ and $m=m_{\mp} \equiv \frac{5-16n+20n^2 \pm 3\sqrt{-1+10n-16n^2+16n^4}}{8(4n-1)(n-1)}$, then there is a solution with $H=H_{\mp}$.

\subsubsection{Summary}

We have seen that anisotropic Kasner-like solutions in higher-order theories of gravity where the Lagrangian depends only on the Kretschmann scalar, $Z \equiv R_{abcd}R^{abcd}$, are possible only if the Lagrangian is of the form $f(Z)=Z^{n}$. Whilst it was claimed in an earlier work by Clifton and Barrow \cite{clifton2} that there are no exact anisotropic solutions of the form (\ref{kas}) in these theories, this is not quite correct. Firstly, for $n=1$, the Kasner solutions of general relativity, for which $H=J=1$, remain valid. This is to be expected, since the theory with $f=\alpha Z$ is a special case of quadratic gravity and all vacuum solutions of general relativity must also be solutions of the pure quadratic theory \cite{hervik2}. However, for all other values of $n$, fully anisotropic solutions of this form with the three Kasner exponents all real and distinct are not possible% only in the special case of quadratic gravity, with $n=1$, when the solutions are subject to the same constraints as in general relativity, with $H=J=1$
. Nevertheless, for choices of $n$ other than $1$, we have seen that solutions which have local rotational symmetry but exhibit an anisotropic third spatial coordinate \emph{are} possible and the Kasner indices are subject to equations (\ref{eq:lrs1}) and (\ref{eq:lrs2}). 

In table \ref{tab:existRiem}, these results are summarised to show whether isotropic power-law, locally rotationally symmetric and fully anisotropic Kasner-like solutions are possible for various choices of the Lagrangian $f(Z)$. Use is made of the definitions given in the previous section; $n_{1} \equiv -\left(\sqrt{5} + \sqrt{3 + 2 \sqrt{5}}\right)/4 \approx -1.24$ and $n_{2}\equiv \left(-\sqrt{5} + \sqrt{3 + 2 \sqrt{5}}\right)/4 \approx 0.12$.

\begin{table}[ht]\small 
\begin{center}
\begin{tabular}[ht]{l||c|c|c|c}
$f(Z)$ & Minkowski & Power-law & LRS Kasner & Anisotropic Kasner \\ \hline\hline
 $Z$ & $\checkmark$ & $\checkmark$ & $\times$ & $\checkmark$ \\
 $Z+\Lambda$ & $\times$ & $\times$ & $\times$ & $\times$ \\
 $Z^{n}, n <n_{1} $ & $\times$ & $\checkmark$ & $\checkmark$ & $\times$ \\
  $Z^{n}, n_{1}<n <0 $ & $\times$ & $\times$ & $\checkmark$ & $\times$ \\
   $Z^{n}, 0<n<n_{2} $ &$\checkmark$ & $\times$ & $\checkmark$& $\times$ \\
  $Z^{n}, n>n_{2}, n\neq 1 $ & $\checkmark$ & $\checkmark$ & $\checkmark$ & $\times$ \\
 $\alpha Z^{n}+\beta Z^{m_{\pm}}, n>0, n\neq \frac{1}{4},1$ & $\checkmark$ & $\checkmark$ & $\times$ & $\times$ 
\end{tabular} \caption{The existence of power-law and Kasner-like vacuum solutions in various models of $f(R_{abcd}R^{abcd})$ theories of gravity.} \label{tab:existRiem}
\end{center}
\end{table}

\subsection{Quadratic gravity; $f=\kappa R+\alpha R^2 +\beta R_{ab}R^{ab} +\gamma R_{abcd}R^{abcd} +\Lambda$ \label{quad}}
The theory of quadratic gravity, in which the Einstein-Hilbert Lagrangian of general relativity is supplemented by quadratic Riemann, Ricci and scalar curvature corrections, is a particularly interesting special case to consider. The Lagrangian is given by \begin{equation}
f=\kappa R+\alpha R^2 +\beta R_{ab}R^{ab} +\gamma R_{abcd}R^{abcd} +\Lambda \, . \label{quadlag} \end{equation} It was shown by Starobinsky \cite{starob1} that addition of quadratic curvature corrections to the Einstein-Hilbert Lagrangian leads to the emergence of inflation. Furthermore, in contrast to general relativity, fourth-order gravity is renormalisable \cite{stelle} and thus it is often motivated as a first-order quantum correction to Einstein's theory. A review of the history of the study of these models of gravity may be found in \cite{schmidt2}. 

Without loss of generality, one may immediately set $\gamma=0$, since the Gauss-Bonnet term, defined by $G \equiv  R^2-4R_{ab}R^{ab}+R_{abcd}R^{abcd}$, is a total divergence in four dimensions, so its variational derivative with respect to the metric does not contribute to the field equations. If the Ricci term is not present in the Lagrangian, i.e. $\beta = 0$, then the theory is a special of the $f(R)$ theories studied earlier and, moreover, it is conformally equivalent to that of a minimally coupled scalar field in general relativity. Thus, one expects that the presence of the Ricci term in the Lagrangian might permit much more diverse anisotropic behaviour, as was found in \cite{hervik}. The quadratic theory is scale-invariant iff $\kappa \Lambda = 0$ and conformally invariant iff $\kappa = \Lambda = 0$ and $3\alpha+\beta =0$.

In order to obtain the correct Newtonian limit in the slow-motion weak-field limit, it is necessary that the fourth-order terms contributed by the quadratic parts of the Lagrangian are exponentially vanishing, rather than oscillatory. To ensure that this is the case, the parameters must satisfy $\beta / \kappa \leq 0$ and $(3\alpha+\beta)/\kappa \geq 0$, with $\kappa \neq 0$.

\subsubsection{Field equations}
For the line element given by (\ref{kas}), the vacuum field equations for the theory of quadratic gravity defined by the quadratic Lagrangian in equation (\ref{quadlag}) are $P^{a}_{b}=0$, and the relevant independent quantities are given by \begin{eqnarray}
P^{a}_{a} &=& \frac{1}{t^{4}}\left((H^{2}-2H + J) (4 (H-3) (3 \alpha + \beta) - 
    \kappa  t^2 ) + 4\Lambda t^{4}\right) \\
P^{0}_{0}    &=& \frac{1}{6t^{4}}\bigl((3 J-H^2)\left(3(J+H^{2})\alpha +(3J-3+2H)\beta \right)+4H^{2}(2H-3)\left(3\alpha+\beta \right)  \nonumber \\ && + 3(J-H^2) \kappa t^2   + 6\Lambda t^4  \bigr) \\
%P^{0}_{0}    &=& \frac{1}{6t^{4}}\bigl(3((3 J-H^2)(J+H^{2})+4H^{2}(2H-3))\alpha +((3 J-H^2)\left(3J-3+2H\right)+4H^{2}(2 H-3))\beta  \nonumber \\ && + 3(J-H^2) \kappa t^2   + 6\Lambda t^4  \bigr) \\
%P^{0}_{0}    &=& \frac{1}{2t^{4}}\bigl((8 H^3 - H^4 + 2 H^2 (J-6) + 3 J^2)\alpha +(2 H^3 + 2 H J + 3 (J-1) J - H^2 (3 + J))\beta  \nonumber \\ && + (J-H^2) \kappa t^2   +2 \Lambda t^4  \bigr) \\
 P^{\mu}_{\mu}- P^{\nu}_{\nu} &=& \frac{1}{t^{4}}(p_{\mu}-p_{\nu})\left((2 (H-3) ((H^{2}-2H + J) \alpha + (J-1) \beta) + (H-1)\kappa t^2 \right) \, \label{eq:qaniso} .
\end{eqnarray} 
%
%\begin{eqnarray}
%P^{0}_{0}    &=& \frac{1}{6t^{4}}\bigl((3 J-H^2)\left(3(J+H^{2})\alpha +(3J-3+2H)\beta \right)+4H^{2}(2H-3)\left(3\alpha+\beta \right)  \nonumber \\ && + 3(J-H^2) \kappa t^2   + 6\Lambda t^4  \bigr) \\
%P^{0}_{0}    &=& \frac{1}{6t^{4}}\bigl(3((3 J-H^2)(J+H^{2})+4H^{2}(2H-3))\alpha +((3 J-H^2)\left(3J-3+2H\right)+4H^{2}(2 H-3))\beta  \nonumber \\ && + 3(J-H^2) \kappa t^2   + 6\Lambda t^4  \bigr) \\
%\end{eqnarray}
As before, $\mu, \nu$ in equation (\ref{eq:qaniso}) for the anisotropic stress are not indices to be summed, but are used there as labels, taking values from $1$ to $3$. Otherwise summation convention is used as normal and overdots represent derivatives with respect to the coordinate time, $t$.

One can immediately see that the cosmological constant, $\Lambda$, must be zero in order for any solutions of this sort to exist.

\subsubsection{$\kappa \neq 0$}
If the Einstein-Hilbert term is present in the Lagrangian, then the only possible isotropic power-law solution is Minkowski space, which requires $\Lambda=0$. There is one family of anistropic solutions, that of general relativity, given by $H=J=1$, provided $\Lambda =0$. In light of the results of the previous sections, this is not unexpected.

\subsubsection{$\kappa = 0$}
In pure quadratic theories, with $\kappa =0$, a non-zero energy-momentum tensor gives rise to a strong gravitational field and consequently spacetime is not asymptotically flat \cite{pech}, but we include them for completeness of this discussion.

Any isotropic solutions of the field equations require $\Lambda =0$ and in general there are two such solutions; Minkowski space and the radiation-like solution with $H=3/2$. For the special case of Weyl gravity, $3\alpha +\beta =0$, \emph{all} power-law isotropic solutions are possible, since in the isotropic case, the contributions to the field equations from the $R^2$ and the Ricci terms in the Lagrangian are the same up to a constant multiple.

There are several possible families of anisotropic solutions; as we have pointed out they all require the cosmological constant, $\Lambda$, to be zero. These solutions and the constraints that must be satisfied are summarised in table \ref{tab:quad}

\begin{table}[ht]\small
%\centering
\begin{center}
\begin{tabular}{ l || p{5.5cm} | p{6.5cm} } 
Class & Solution constraints & Validity \\[2pt] \hline \hline 
\textbf{I} & $H=J=1$ & This is a vacuum solution for all values of $\alpha, \beta$ and $\kappa$. \\[4pt]
\textbf{II} & $J=\frac{1}{2}(3-2H+H^2)$ & These are solutions for the Weyl theory of gravity only, that is if $3\alpha+\beta=0=\kappa$. \\[4pt]
\textbf{III} & $H=3$, \newline $J=\frac{1}{\alpha+\beta}(-3\alpha+\beta-2\sqrt{-2\beta(3\alpha+\beta)})$ & These are valid with real exponents if $\kappa = 0$ and $-\beta /3 < \alpha < -\beta$. \\[4pt]
\textbf{IV} & $H=3$, \newline $J=\frac{1}{\alpha+\beta}(-3\alpha+\beta+2\sqrt{-2\beta(3\alpha+\beta)})$ & These are valid with real exponents if $\kappa = 0$ and $-\beta  < \alpha < -\beta /3$. \\[4pt]
\end{tabular}
\caption{Kasner-like vacuum solutions in pure quadratic gravity.} \label{tab:quad}
\end{center}
\end{table}

Recall that the first class of solutions are vacuum solutions in general relativity and so have vanishing Ricci tensor. As a consequence, they must also be vacuum solutions of the quadratic theory for all values of the parameters $\alpha, \beta$ and $ \kappa$. Indeed, we have already seen that they are solutions for the special cases of quadratic Lagrangians within the more general models we have previously investigated. The constraint on the solutions of class II implies that $p_{3}= p_{1}+p_{2}-1 \pm 2 \sqrt{(p_{1}-1)(p_{2}-1)}$, and so all three Kasner exponents may be positive and greater than unity. Classes I-III were previously found by Deruelle \cite{deruelle}.

\subsection{Gauss-Bonnet theories; $f=f(G), f=R+\hat{f}(G)$}
In four dimensions, the Gauss-Bonnet term is a topological invariant and its variation does not contribute to the field equations, though it may give rise to interesting cosmological effects in higher dimensions \cite{nojiri1}. These terms arise in the low energy effective actions of string theory, however modified Gauss-Bonnet theories have also been proposed as a form of gravitational dark energy capable of successfully describing cosmology at late times \cite{nojiri2}. Here, we shall consider the situation where the Lagrangian is a general function of the Gauss-Bonnet invariant, i.e. $f=f(G)$ and also the case $f=R+\hat{f}(G)$, where the Gauss-Bonnet term is defined by $G\equiv R^2-4R_{ab}R^{ab}+R_{abcd}R^{abcd}$, and takes the form \begin{equation}
G=\frac{4p_{1}}{t^4} (H-3) (2 p_{1}\!^2 - 2 H p_{1} + H^2 - J)
\end{equation} for the metric given by (\ref{kas}). 

The contributions $\hat{P}^a_b$ to the field equations due to a function $f(G)$ in the Lagrangian are given by \begin{eqnarray}
\hat{P}^{0}_{0} &=& -\frac{1}{2}f + 4\frac{p_{1}p_{2}p_{3} }{t^4}\left((H-3) f_G - 3 t \dot{f}_{G} \right) \: ,\\
\hat{P}^{i}_{i} &=& -2f +\frac{2}{t^4}(8 p_{1}p_{2}p_{3} (H-3) f_G + (J-H^2) ((H-2)t \dot{f}_{G} + t^2 \ddot{f}_{G}) \: , \\
\hat{P}^{\mu}_{\mu}-\hat{P}^{\nu}_{\nu} &=& \frac{4}{t^3}\left(p_{\mu}-p_{\nu}\right)\left(H-p_{\mu}-p_{\nu}\right)\left((H-2) \dot{f}_G + t \ddot{f}_{G} \right) \: , \label{eq:gbaniso}
\end{eqnarray} where once again, $\mu, \nu$ in equation (\ref{eq:gbaniso}) are not indices to be summed, but labels taking values from $1$ to $3$. Otherwise summation convention is used as normal. Here, $f_{G}$ is used to denote $ \frac{df}{dG}$ and overdots represent derivatives with respect to the time coordinate $t$.

The anisotropic stress will vanish independently of $f$ both for isotropic solutions, ie $p_{1}=p_{2}=p_{3}$, and solutions in which two of the Kasner exponents are zero, and also in theories in which the Lagrangian is a power of the Gauss-Bonnet term, provided that $H$ is suitably chosen in this case.

\subsubsection{$f=f(G)$}
Except for the trivial case of $f(G)=G$, the only possible real anisotropic solutions have $G=0$. Furthermore, for any Kasner solutions to be possible, it is required that $f(0)=0$. Provided $f_{G}(0)$ does not diverge, then any solution of $G=0$ defines a two-parameter family of exact Kasner-like solutions. If, on the other hand, $f_{G}(0)$ is divergent, then for general such $f(G)$, there is a one-parameter family of solutions where the three Kasner indices are given by $p_{1}=p_{2}=0$, with the third index free, plus permuations. In the special case $f(G)=\alpha G \log{G}$, there are additional families of solutions given by $p_{1}=0$, with $p_{2}$ and $p_{3}$ free (and permutations), or $J=2 p_{1}\!^2 - 2 H p_{1} + H^2$.

\subsubsection{$f=R+\hat{f}(G)$}
The only real solutions with $G=0$ and $R_{ab}=0$ are Minkowski space and the Milne model, corresponding to $p_{1}=1, p_{2}=p_{3}=0$ (plus permutations), which are vacuum solutions whenever $\hat{f}(0)=0$. Consequently, only these solutions can separately satisfy both the vacuum Einstein equations, $G_{ab}=0$, and the vacuum equations due to the purely Gauss-Bonnet terms, $\hat{P}_{ab}=0$. Other solutions must have $G \neq 0$ and the Einstein-Hilbert terms must balance the Gauss-Bonnet terms, that is to say they must be of the same order in time. Therefore, the possible real Kasner-like vacuum solutions in these theories may be summarised as in table \ref{tab:gb}.

\begin{table}[ht]\small 
\begin{center}   
 \begin{tabular}{ l || p{6cm} | p{6cm} } 
Class & Solution constraints & Validity \\[2pt] \hline \hline 
\textbf{I} & $p_{1}=p_{2}=p_{3}=-1$ & This is a vacuum solution if $f=\alpha \sqrt {G}+\frac{\sqrt{3G}}{2} \log {G}$. \\[4pt]
\textbf{II} & $p_{1}=p_{2}=p_{3}=-3$ & This is a vacuum solution if $f=\alpha G \log{G}- 3\sqrt{2G}$. \\[4pt]
\textbf{III} & $p_{1}=p_{2}=p_{3}=\frac{1}{6-\alpha^2}(3+\alpha^2 \pm \sqrt{12\alpha^2+9})$ & This is a vacuum solution if $f=\alpha \sqrt{G}$. \\[4pt]
\textbf{IV} & $p_{1}=1, p_{2}=p_{3}=1+\alpha^2$ \newline (and permutations) & These are vacuum solutions if $f=\alpha \sqrt{G}$. \\[4pt]
\textbf{V} & $H=1, J=1 + 4 p_{1}\alpha^2 + 4p_{1}\alpha \sqrt{1-p_{1}+\alpha^2}$ & These are vacuum solutions if $f=\alpha \sqrt{G}$. \\[4pt]
\textbf{VI} &$H=1, J=1 + 4 p_{1}\alpha^2 - 4p_{1}\alpha \sqrt{1-p_{1}+\alpha^2}$ & These are vacuum solutions if $f=\alpha \sqrt{G}$. \\[4pt]
\end{tabular} \caption{Power-law and Kasner-like vacuum solutions in a modified Gauss-Bonnet gravity.} \label{tab:gb}
\end{center}
\end{table}

We can see that although there are several different and interesting types of Kasner-like solutions in these theories, they are only valid for a small set of Lagrangians. This is because only the Minkowski and Milne universes solve both the vacuum Einstein equations, $G_{ab}=0$, and the vacuum field equations due to the purely Gauss-Bonnet terms in the action. Thus, terms in the field equations due to the Einstein-Hilbert term must be balanced by those due to $\hat{f}(G)$ and so these terms must be of the same order in time.

\subsection{Weyl theories of gravity}
We now consider the situation where the Lagrangian is a general function of the Weyl invariant, i.e. $f=f(W)$, where the Weyl term is defined by $W\equiv \frac{1}{3}R^2-2R_{ab}R^{ab}+R_{abcd}R^{abcd}$. We may write \begin{equation*}
W=\frac{4}{3t^{4}}\left(  (p_{1} - p_{3}) (p_{1} - p_{2})(p_{1}-1)^2  -(p_{2}-p_{3})(p_{1}-p_{2})(p_{2}-1)^2+(p_{1} - p_{3}) (p_{2}-p_{3})(p_{3} - 1)^2 \right) \end{equation*} If we assume, without loss of generality, $p_{1}\geq p_{2} \geq p_{3}$, then we see that the first and third terms are positive, but the second is negative. However, the middle term must be no greater in absolute magnitude than the third term if $p_{2}<1$ and similarly it must be no greater in magnitude than the first term if $p_{2}\geq1$. Thus the Weyl tensor is non-negative and is zero only if each term vanishes separately. This requires that either the metric is locally rotationally symmetric with wlog $p_{2}=p_{3}$ and $p_{1}=1$, or it is isotropic, $p_{1}=p_{2}=p_{3}$. The vacuum field equations are $P^{a}_{b}=0$, with \begin{eqnarray}
P^{i}_{i} &=& -2f + 2W f_{W} \: , \\
P^{\alpha}_{\alpha} &=& -\frac{3}{2} f + 
 \frac{2}{3t^4} (18 p_{1}p_{2}p_{3} + H^2 - 2 H^3 - 3 J + 4 H J) ((H-3) f_{W} + t \dot{f}_{W})) \: ,\\
\hat{P}^{\mu}_{\mu}-\hat{P}^{\nu}_{\nu} &=& \frac{1}{3t^4}\left(p_{\mu}-p_{\nu}\right)\biggl(12t (H-p_{\mu}-p_{\nu}) ((H-2) 
\dot{f}_{W} + t 
\ddot{f}_{W})  + 
 4 (H - 3) (2 (J - 1) - (H - 1)^2) f_{W} \nonumber \\ && + 
 2 (3 (H - 1)- 5 (H - 1)^2 + 4 (J - 1) ) t  
\dot{f}_{W} - 6 (H - 1) t^2 
\ddot{f}_{W} \biggr) \: , \label{eq:weylaniso}
\end{eqnarray} where once again, $\mu, \nu$ in equation (\ref{eq:gbaniso}) are not indices to be summed, but labels taking values from $1$ to $3$. Otherwise summation convention is used as normal, $f_{W}$ is used to denote $f'(W) \equiv \frac{df}{dW}$ and overdots represent derivatives with respect to the time coordinate $t$. The equations (\ref{eq:weylaniso}) for the anisotropic stress may be combined to obtain \begin{equation}
%\left(p_{1}-p_{2}\right)\left(P^{2}_{2}-P^{3}_{3}\right)-\left(p_{2}-p_{3}\right)\left(P^{1}_{1}-P^{2}_{2}\right) \\
%&=&
0=
 \frac{4}{t^3}\left(p_{1}-p_{2}\right)\left(p_{1}-p_{3}\right)\left(p_{2}-p_{3}\right) \left((H-2) \dot{f}_{W} + t \ddot{f}_{W}\right) \: .
\end{equation}

For fully anisotropic solutions, the Weyl tensor is non-zero, and as before one can treat the field equations as differential equations for $f$, solving them to find the required form of Lagrangian for the higher-order theory to possess such solutions. It is found that the only possible choice is that of Weyl gravity, $f(W) = \alpha W$, and that solutions require $p_{3}=p_{1}+p_{2}-1 \pm 2\sqrt{(p_{1}-1)(p_{2}-1)}$; this is a special case of the quadratic Lagrangians considered before in section \ref{quad}, corresponding to $3\alpha+\beta=0$.

For solutions of the vacuum field equations where the Weyl tensor vanishes, it is necessary that $f(0) = 0$. In fact, if $f(0)=0$ and $f'(W)$ does not diverge at $W=0$, then all solutions of $W=0$ are solutions of the vacuum field equations. However, if $f(0)=0$ and $f'(0)$ diverges then we must consider both the isotropic and locally rotationally symmetric types of solution in turn as the existence of such solutions will depend upon the manner of this divergence.

These results are summarised in table \ref{tab:weyl}.

\begin{table}[ht]\small 
\begin{center}   
 \begin{tabular}{ l || p{5.5cm} | p{7cm} } 
Class & Solution constraints & Validity \\[2pt] \hline \hline 
\textbf{I} & $p_{1}=p_{2}=p_{3}$ & These are vacuum solutions for all values of $p_{1}$ if $f(0)=0$ and either $f'(0)$ converges or diverges due to terms in $f(W)$ of the form $W^{n}$ with $1/2 < n < 1$. \\[4pt]
\textbf{IIa} & $p_{1}=p_{2}=p_{3}=2n-1$ & This subset of the class I solutions remains a vacuum solution if $f(0)=0$ and the divergence in $f'(0)$ is due to terms in $f(W)$ of the form $W^{n}$ with $0 < n < 1/2$.
 \\[4pt]
 \textbf{IIb} & $p_{1}=p_{2}=p_{3}=(4n-1)/3$ & This subset of the class I solutions remains a vacuum solution if $f(0)=0$ and the divergence in $f'(0)$ is due to terms in $f(W)$ of the form $W^{n}$ with $0 < n < 1/2$.
 \\[4pt]
 \textbf{III} & $p_{1}=1, p_{2}=p_{3}$ \newline (and permutations) & These are vacuum solutions if $f(0)=0$ and $f'(0)$ either converges or diverges due to terms in $f(W)$ of the form $W^{n}$ with $1/2 < n < 1$. \\[4pt]
\textbf{IV} & $p_{1}=1, p_{2}=p_{3}=2n-1$ \newline (and permutations) & This subset of the class III solutions remains a vacuum solution if $f(0)=0$ and the divergence in $f'(0)$ is due to terms in $f(W)$ of the form $W^{n}$ with $0 < n < 1/2$.
 \\[4pt]
 \textbf{V} & $p_{3}=p_{1}+p_{2}-1 \pm 2\sqrt{(p_{1}-1)(p_{2}-1)}$ & This is a vacuum solution in Weyl gravity, $f(W)=\alpha W$.
 \\[4pt]
\end{tabular} \caption{Power-law and Kasner-like vacuum solutions in Weyl gravity.} \label{tab:weyl}
\end{center}
\end{table}
Note that if $f(W)$ contains a term $\sqrt{W}$, i.e. $n=1/2$, then the only solution is that of the Milne model, with metric given by
\begin{equation*}
 ds^2 = -dt^2+t^2 dx^2 + dy^2 + dz^2 \: ,
\end{equation*} which we have seen is Minkowski space in a different coordinate system. Finally we point out that if the $f'(W)$ diverges at $W=0$ due to a type of term other than a power law, there are no vacuum Kasner-like solutions in that theory.

\subsection{Homogeneous Lagrangians \label{sec:homogeneous}}
Thus far, this study has considered models in which the Lagrangian is a general function of one of the curvature scalars, $X\equiv R$, $Y \equiv R^{ab}R_{ab}$, or $Z \equiv R^{abcd}R_{abcd}$, or of a particular combination of these. Consequently, in each case there has been only one timescale in the problem, and, by substituting for the time variable using the appropriate curvature invariant, it has proved possible to derive all possible solutions and their existence conditions within these wide classes of models.

However, it is useful to also consider more general Lagrangians depending on more than just one variable. We may also conisder Lagrangians which are homogeneous in $X^{2}, Y$ and $Z$, that is to say for all values of $\lambda$, $f(\lambda X^2, \lambda Y, \lambda Z) = \lambda^{n}f(X^2, Y, Z)$ for some $n$. $X^{2}$ is chosen as the argument here, rather than $X$, in order that each term in the function will be of the same order in time. Thus, functions of this form will be relevant where the dominant terms at early or late times are more complicated than in those Lagrangians studied previously, such as a monomial in $X,Y,Z$. We may write 
\begin{equation}
f\left(X,Y,Z\right)= X^{2n} \alpha \left(\theta, \phi \right)  \label{homogeneous},
\end{equation} where $\alpha$ is a general differentiable function of $\theta \equiv \frac{X^{2}}{Z}$ and $ \phi \equiv \frac{Y}{Z}$, and further define \begin{eqnarray}
\beta & \equiv & \frac{\partial \alpha}{\partial \theta} \: ,\\
\gamma & \equiv & \frac{\partial \alpha}{\partial \phi}  \: . \end{eqnarray} 
%where $\theta \equiv \frac{X^{2}}{Z}, \phi \equiv \frac{Y}{Z}$. 
It is important to note that $\alpha, \beta$ and $\gamma$ take constant values which are dependent upon the exact form of the Lagrangian $f$ and also the values of $\theta$ and $\phi$ for a particular solution. For example, in the case of the monomial given by \begin{equation}
f=\xi X^{\sigma} Y^{\mu} Z^{\nu} ,
\end{equation} we have
\begin{eqnarray}
2n &=& \sigma+2\mu+2\nu \: , \\
\alpha &=& \xi \left(\frac{X^{2}}{Z}\right)^{-(\mu+\nu)}\left(\frac{Y}{Z}\right)^{\mu} \: , \\
\beta &=& -(\mu+\nu)\frac{Z}{X^{2}} \alpha \: , \\
\gamma &=& \mu \frac{Z}{Y} \alpha \: .
\end{eqnarray}

Inserting the general form (\ref{homogeneous}) for the homogeneous Lagrangian into the vacuum field equations for the metric (\ref{kas}), one obtains the equation
\begin{eqnarray}
0 &=& \left(p_{2}-p_{3}\right)\left(P^{1}_{1}-P^{3}_{3}\right)-\left(p_{1}-p_{3}\right)\left(P^{2}_{2}-P^{3}_{3}\right) \nonumber \\
&=&
\frac{16 t^{4} X^{2n}}{Z^{2}}\left(p_{1}-p_{2} \right)\left(p_{1}-p_{3}\right)\left(p_{2}-p_{3}\right)(n-1)(1+H-4n)\left(\beta X^{2}+\gamma Y \right) \: . \label{eq:aniso}
\end{eqnarray}
We will consider in turn the solutions of this equation in the remaining field equations.

\subsubsection{Isotropic solutions}
In the isotropic case, $p_{1}=p_{2}=p_{3}$, there is only one independent field equation, given by
\begin{eqnarray*}0 & = & p_{1}\left(p_{1}(2p_{1}-1)\right)^{2n-1}\bigl(\alpha(2p_{1}^2-2p_{1}+1)^2(1+2p_{1}(n-1)-6n+8n^2)\\
& & -(6\beta+\gamma) p_{1}(1-2p_{1})^2(1+3p_{1}-4n)\bigr) \: .
\end{eqnarray*}

Recall that $\alpha, \beta$ and $\gamma$ are constants which will depend upon $p_{1}$ according to the exact form of the Lagrangian $f$. Thus, given a choice of $f$, this equation provides a necessary and sufficient condition for there to be a isotropic power-law solution expanding as $t^{p_{1}}$.

\subsubsection{Locally rotationally symmetric solutions}
In the case of locally rotationally symmetric (LRS) solutions, there are two independent field equations. Without loss of generality, we set $p_{2}=p_{3} \neq p_{1}$. 

LRS solutions with $X=0$ are a special case of the fully anisotropic ones of that kind and will satisfy the same existence conditions. Similarly, if $p_{1}+2p_{2}=4n-1$, then there are LRS solutions provided $\alpha X^{2n}=0$ can be solved simultaneously. This requires either the additional constraint $\alpha=0$ to be satisfied, or if $1/4<n<5/8$ then there are real solutions of this form with $X=0$.  Both these types of LRS solution satisfy the same constraints as in the fully anisotropic case, and so these will be discussed at greater length in the next section.

If we define $\hat{X}\equiv t^2 X, \hat{Y}\equiv t^4 Y$ and $\hat{Z} \equiv t^4 Z$ for convenience, then, for $X\neq 0$, these may be reduced to
\begin{eqnarray*}
0 &=& \biggl(n \alpha  \hat{Z}^2 - 2 \hat{X} (2-3p_{1}-2p_{2} +(p_{1}-p_{2})^2 + 4 n(p_{1}+p_{2}-1)) (\beta \hat{X}^2 + \gamma \hat{Y}) + \beta \hat{X}^2 \hat{Z} \\ 
&& + \gamma \hat{X}\hat{Z} (p_{1}\!^2 + 2 p_{2}\!^2 - 1 + 2(p_{1}+2p_{2}-1)(n-1)) \biggr)(1+p_{1}+2p_{2}-4n) \: , \\
0&=& \alpha \hat{Z}^2\left(4n \left(p_{1}\!^2 + 2p_{2}\!^2+ (4n-3)(p_{1}+2p_{2})\right) - \hat{X} \right) \\ 
&& +16\hat{X} p_{2}\!^2(1+p_{1}+2p_{2}-4n)  \biggl(\beta \hat{X}\left(p_{1}\!^3 -2p_{1}\!^2p_{2}+ p_{1}p_{2}\!^2 -5p_{1}\!^2-2p_{1}p_{2} - 2p_{2}\!^2\right) + \gamma \bigl( (2-3p_{2}) p_{2}\!^3 \\
&& +p_{1}\!^5 + 2 p_{1} (p_{2}-1) p_{2}\!^3 - 2 p_{1}\!^4 (1 + p_{2}) + 
   p_{1}\!^2 p_{2} (6 - 7 p_{2} - 4 p_{2}\!^2) + p_{1}\!^3 (1 - 4 p_{2} + 3 p_{2}\!^2)\bigr)\biggr) \: .
\end{eqnarray*}

In this way, the field equations give relationships between the theory-dependent constants $\alpha, \beta$ and $\gamma$ which must be satisfied if the theory is to contain any other LRS solutions.

\subsubsection{Anisotropic solutions}

We now consider those solutions of equation (\ref{eq:aniso}) which are fully anisotropic with $p_{\alpha}$ all distinct. We have seen that for metrics described by the line element (\ref{kas}), $Z=0$ only for Minkowski space and the Milne model. Thus for fully anisotropic solutions, $Z$ will be non-zero. The only real Kasner-like solutions with $X=Y=0$ are those of general relativity, which satisfy $H=J=1$. Since the leading order terms in the field equations as $X,Y \rightarrow 0$ behave like $\alpha(2n-1)X^{2n-1}$, then if $\alpha$ diverges at $X=0$, faster than $X^{1-2n}$, there will be no solutions. However, if $\alpha$ is finite at $Y=0$, one either requires that $n \geq 1/2$ or that $\alpha=0$ and $n>0$ so that the next leading order terms do not diverge. For solutions with $X=0$ but $Y\neq 0$, the leading order terms in the field equations as $X \rightarrow 0$ behave like $\alpha(H-4n+1) X^{2n-1}$, so if $\alpha$ remains finite at $X=0$, one either requires that $n \geq 1/2$  or that either $H=4n-1$ or $\alpha=0$ and $n>0$ so that the next leading order terms do not diverge.

If there are configurations of the Kasner exponents such that $\alpha=\beta=\gamma=0$, then these will be solutions of the theory. Recall that $\alpha, \beta$ and $\gamma$ depend on the explicit form of the Lagrangian.

Substituting $H=4n-1$ into the field equations leads to the necessary and sufficient condition $\alpha X^{2n}=0$. For $X=0$ with $H=4n-1$, this implies that $J=16n(1-n)-3$ and requires that $1/4 \leq n \leq 5/8$ for these solutions to be real. Again, since $\alpha$ depends on the explicit form of the Lagrangian and the values of the Kasner exponents and so $\alpha=0$ defines an extra constraint on the solution.

In addition to those solutions which are a special case of the ones with $H=4n-1$ and $\alpha=0$, there is another class of solutions if $n=1$. This class of solutions exists if
\begin{eqnarray} \alpha &=&-\frac{\gamma}{4 Z}\left(X^2-4Y+Z\right)  \quad \text {and}\\
\beta &=&-\frac{\gamma}{4 X^2}\left(4Y-Z\right) \: . \end{eqnarray}
Again, $\alpha, \beta$ and $\gamma$ will depend on the values of the Kasner exponents through the curvature scalars $X, Y$ and $Z$ and so these equations will typically give two further constraints on the possible set of solutions. However, if for example, $\alpha=\alpha(X^{2}/(4Y-Z))$, then the second of these equations will be trivial and if $\alpha=\lambda(X^2-4Y+Z)/X^{2}$ for some constant $\lambda$, then both of these will be trivial.

Finally, there are several additional solutions if $\beta X^2 + \gamma Y=0$. This equation is satisfied trivially if the Lagrangian $f$ is independent of $Z$. Otherwise, it provides a constraint on the solutions. The remaining field equations then show that if $n=1$ there are solutions subject to $H=1$ and \begin{equation}
\alpha=\frac{4\beta}{Z t^{4}}(1-J) \: .
\end{equation} If $n=1/2$, there are solutions subject to \begin{equation}
\alpha = 2\beta \frac{X^{2}}{t^{4}Y Z} (1-H) (H^2 - J) \: . \label{nhalf} \end{equation} Finally, for general $n$ there are solutions of this sort if \begin{eqnarray}
\alpha &=& 2\beta \frac{X^2}{n t^{4}Y Z}  (H-1) \left(H-H^2 +n\left(H^{2}-2H+J \right)\right) \quad \text{and}\\ 
J &=& \frac{1}{2n}\left(1 - 2 H + H^2 - 6 n + 8 H n + 8 n^2 - 8 H n^2\right)  \: .
\end{eqnarray}

\subsubsection{An explicit example}
As an explicit example, we shall consider fully anisotropic Kasner-like solutions in the model of Br\"{u}ning, Coule and Xu \cite{coule}, which is described by the Lagrangian \begin{equation}
%f = R +\lambda \frac{R_{ab}R^{ab}}{R}+\tau \frac{R_{abcd}R^{abcd}}{R} 
f = X +\lambda \frac{Y}{X}+\tau \frac{Z}{X}
\: , \label{f:coule} \end{equation} where $\lambda$ and $\tau$ are constants, and was studied there in the context of FRW solutions. Note that the model reduces to general relativity in the limit  $\lambda=\tau=0$, so in the discussion that follows we will assume that $\lambda$ and $\tau$ are not both zero. In terms of our notation, 
\begin{eqnarray*}
% n &=& \frac{1}{2} \: ,\\
% \alpha &=& 1+\lambda \frac{R_{ab}R^{ab}}{R^{2}}+\tau \frac{R_{abcd}R^{abcd}}{R^{2}} \: ,\\ 
% \beta &=& -\frac{R_{abcd}R^{abcd}}{R^{4}}\left(\lambda R_{ab}R^{ab}+\tau R_{abcd}R^{abcd}\right) \quad \text{and} \\
% \gamma &=& \lambda \frac{R_{abcd}R^{abcd}}{R^{2}} \: .
n &=& \frac{1}{2} \: ,\\
\alpha &=& 1+\lambda \frac{Y}{X^{2}}+\tau \frac{Z}{X^{2}} \: ,\\ 
\beta &=& -\frac{Z}{X^{4}}\left(\lambda Y+\tau Z \right) \quad \text{and} \\
\gamma &=& \lambda \frac{Z}{X^{2}} \: .
\end{eqnarray*} Thus we can explicitly observe the dependence of $\alpha, \beta$ and $\gamma$ upon the values of the Kasner parameters $p_{\alpha}$ and in particular for this model we note that $\alpha$ diverges for the general relativistic solution, $H=J=1$. 

For anisotropic solutions in this model, with the Kasner exponents $p_{\alpha}$ all distinct, equation (\ref{eq:aniso}) reduces to \begin{equation}
0=\tau \left(\frac{H-1}{H^2-2H+J}\right) \: ,
\end{equation}and so the possible solutions depend on the value of $\tau$. For the special case when $\tau$ is zero, the field equations reduce to the single equation \begin{equation}
\frac{2 H^3 -3H^2+ J + 2 H J -H^{2}J- J^2 }{(H^2-2H+J)^{2}}\lambda = 1 \: .
\end{equation} This equation allows a two-parameter family of solutions to be found with real-valued Kasner exponents for all values of $\lambda$ except those in the range $-1<\lambda < 0$.

For the general case with $\tau$ non-zero, solutions require $H=1$ and 
% \begin{equation}
% (1 + \lambda + 3 \tau)(J-1)^2 +8 \tau p_{1}(J-1) - 16 \tau  p_{1}\!^{2} (p_{1} - 1)=0 \: . \end{equation}
% The solutions of this have real values of $J$, provided $a$ can be found such that \begin{equation}
% 1+\frac{1}{\tau}(a-1)(1+\lambda+3\tau) \geq 0 \: .
% \end{equation} 
% The solutions of this have real values of $J$, provided $a$ can be found such that \begin{equation}
% 1+\frac{1}{\tau}(a-1)(1+\lambda+3\tau) \geq 0 \: .
% \end{equation} 
\begin{equation}
\xi(J-1)^2 +8 p_{1}(J-1) - 16  p_{1}\!^{2} (p_{1} - 1)=0 \: ,\label{ineq1} \end{equation}
where we have defined $\xi \equiv (1+3\lambda+\tau)/\tau$. These solutions therefore are described by the single free paramter $p_1$. Seeking solutions for which the Kasner exponents $p_{\alpha}$ take real values, it is necessary to choose $p_{1}$ such that \begin{equation}
1+\xi(p_{1}-1) \geq  0 \end{equation} and either \begin{eqnarray}
3p_{1}\!^{2}+2\left(\frac{4}{\xi}-1\right)p_{1}-1 &<& 0 \qquad \text{or} \label{ineq2}\\
\xi\left(\xi(1+3p_{1})^{2}-16p_{1}\right) & < & 0 \: . \label{ineq3}
\end{eqnarray}It is possible to find $p_{1}$ such that the inequalities (\ref{ineq1}) and (\ref{ineq2}) may be satisfied simultaneously if $\xi <5/4$, whilst the inequalities (\ref{ineq1}) and (\ref{ineq3}) may be satisfied simultaneously provided $\xi <4/3$. Thus, this bound provides the restriction on the allowable choices of $\lambda$ and $\tau$ such that the general model (\ref{f:coule}) will contain anisotropic Kasner-like solutions of the form (\ref{kas}).

If we compare these findings with those in the previous section, when the general case of a homogeneous Lagrangian was discussed, we see that these correspond to the solutions of equation (\ref{nhalf}). When $\tau=0$, the equation $\beta X^2 + \gamma Y=0$ is trivial, but for non-zero $\tau$, it provides an additional constraint and reduces the number of free parameters in the solution from two to one. There are no anisotropic solutions corresponding to $X=0$, since (except in the limiting case of general relativity) $\alpha X$ diverges there.

\subsubsection{Remarks}
In this section, we have studied those Lagrangians which are homogeneous in time for the metrics (\ref{kas}), and which may be written in the form (\ref{homogeneous}). We have found the existence conditions for all possible Kasner-like solutions in this class of theories. Although it is impossible to find the exact forms of these solutions without explicitly defining the Lagrangian, and the dependence of the functionals $\alpha, \beta$ and $\gamma$ on the Kasner parameters may be complicated, we have developed a general framework which can be used to obtain all possible solutions once the gravitational theory has been defined. This framework was then explicitly demonstrated for the model of Br\"{u}ning, Coule and Xu \cite{coule}.
 
Since the terms in the field equations must vanish at each order in time, then unless there exists a particular configuration of Kasner exponents such that the Lagrangian $f(X,Y,Z)$ is a constant, it is reasonable to expect that the solutions found in this section will be the \emph{only} Kasner-like solutions with real Kasner indices that are possible in higher-order metric theories of gravity derived from Lagrangians with the general form $f(R, R_{ab}R^{ab}, R_{abcd}R^{abcd})$.

\section{Anisotropically Inflating Solutions}
In the preceeding sections, we have studied higher-order theories of gravity in which the Lagrangian is dependent upon the scalar curvature, $X \equiv R$, the Ricci invariant, $Y \equiv R_{ab}R^{ab}$, and the Kretschmann scalar, $Z \equiv R_{abcd}R^{abcd}$, to find the conditions required on the form of the Lagrangian for a particular theory to contain Kasner-like vacuum solutions, where the line element is described by (\ref{kas}). Within the context of this class of higher-order gravitational theories, it is interesting to also study the possibility of other simple anisotropic exact solutions. A natural yet simple extension is to consider a Bianchi type I solution which describes exponential but anisotropic expansion, where the metric is of an anisotropic deSitter-like form;
\begin{equation}
 ds^2=-dt^2+e^{2p_{1}t} dx^2 +e^{2p_{2}t} dy^2 +e^{2p_{3}t} dz^2 , \label{ds}
\end{equation}
and the exponents $p_{\alpha}$ are real. 

In this section, we shall investigate whether such metrics solve the vacuum field equations of these theories and we include also the possibility of a cosmological constant. Such solutions were found to be possible in theories with quadratic Lagrangians, provided that the Ricci term is present \cite{hervik} and their existence demonstrates that the cosmic no-hair theorem of general relativity, which states that the presence of a positive cosmological constant drives the solution towards the deSitter one at late times, cannot be extended to higher-order theories in general.

The three curvature scalars $X,Y,Z$ are constants, and take the values
\begin{eqnarray*}
X &=& H^2+J \: , \\
Y &=& (H^2+J)J \: ,\\
Z &=& 2(J^2+M) \: ,%= 3J^2+2H^2J-H^4+8H abc,
\end{eqnarray*}
where the useful definitions $H \equiv p_{1}+p_{2}+p_{3}, J \equiv p_{1}\!^2+p_{2}\!^2+p_{3}\!^2$ and $M \equiv p_{1}\!^4+p_{2}\!^4+p_{3}\!^4$ have been made. Note that all three curvature scalars $X,Y$ and $Z$ are constant and also positive unless $p_{1}=p_{2}=p_{3}=0$, corresponding to Minkowski space. Thus, unlike the situation for the Kasner solutions discussed previously, it is not necessary for example that $R=0$ for $f(R)$ to be zero. Furthermore, since the curvature scalars are constant, the field equations simplify substantially \cite{clifton4}, and for the isotropic case they reduce to the single equation
\begin{equation}
\frac{1}{2} f = \frac{H^2}{3} f_{X}+\frac{2}{9}H^4 f_{Y} +\frac{4}{27}H^4 f_{Z} \: ,
\end{equation} which must be satisfied by $f$ if the de Sitter universe is to be a solution in a particular higher-order theory of gravity of the form $f(X,Y,Z)$.

For all vacuum \emph{anisotropic} solutions of the form (\ref{ds}), they become
\begin{eqnarray}
f &=& 8H p_{1}p_{2}p_{3} f_{Z} \: ,  \label{eq:ds1} \\
f_X+ 2J f_Y &=& 2(H^2-3J) f_Z \: , \label{eq:ds2}
\end{eqnarray}
where, as before, subscripts are used to denote differentiation of $f$ with respect to that curvature scalar, so that, for example, $f_{X} \equiv \frac{\partial f}{\partial X}$. Any cosmological constant is to be included in the function $f$.

If the Lagrangian is a function of only one of the three curvature invariants, $f=f(\xi)$ say, where $\xi \in \{ X, Y, Z \}$, then the field equations for anisotropic solutions simplify further to
\begin{eqnarray}
f &=& 0 \: , \\
f_{\xi} &=& 0 \: .
\end{eqnarray}
Thus the theory will contain a two-parameter family of exact anisotropic deSitter-like solutions if the Lagrangian, $f$, has a \textit{positive} double root, $\xi=\xi_{0}>0$, the simplest example of which being a quadratic Lagrangian, $f=\alpha(\xi-\xi_{0})^2$. It should be noted that for the case where $\xi \equiv R$, the scalar curvature, choosing the sign of the Lagrangian to give the correct Newtonian limit would give a negative cosmological constant due to the required positivity of $R_0$. 

In this way, some degree of fine-tuning is required in the Lagrangian for such solutions to exist. However, more general types of model do not require such fine-tuning for solutions of this sort to exist. In particular, if the Lagrangian is of the form $f=f(Y/X^{2})$, then equation (\ref{eq:ds2}) is identically satisfied and so it is only required that there exist positive roots of $f=0$.

%\subsection{Kasner-like dust solutions}
%Here, we consider solutions in $f=f(R)$ theories of the form
%\begin{equation}
% ds^2=-dt^2+\Sigma_{i=1}^{3} a_i(t) ^2 (dx^i)^2 \, , \label{dust}
%\end{equation}
%where $a_i = t^{\alpha_i}(t-t_0)^{m-\alpha_i}, H \equiv \Sigma_{i=1}^{3} \alpha_i , J \equiv \Sigma_{i=1}^{3} (\alpha_i)^2$, in the presence of a perfect fluid with barotropic equation of state $p=w \rho$.
%
%The only possible anisotropic solutions are :
%\begin{itemize}
%\item $m=2/3, w=0, H=J=1, f=\alpha R, \rho = \rho_0 / [t(t-t_0)], \rho_0=4\alpha /3, R=4/ 3[t(t-t_0)] $ (This is the general relativistic solution).
%\item $m=1, w=0, H=3/2, J=3/4, f=\alpha R^{(3/2)} , \rho = \rho_0 / [t(t-t_0)]^{(3/2)}, \rho_0=-3\alpha \sqrt{6}/2, R=6/ [t(t-t_0)] $
%\item $p=\rho=0, m=1/2, J=2H-H^2, f=f_R=0 , R=0 $
%\item $p=\rho=0, f=0 $ Minkowski space.
%\end{itemize}
%There may be some isotropic solutions, $a_i = t^{H/3}(t-t_0)^{M}$. NB Need to consider dusty Kasner model, $\rho = \rho_0 t^{-H}$. in earlier analysis etc.

\section{Conclusions}
We have studied some aniosotropic cosmological Bianchi type I solutions to a wide class of higher-order theories of gravity derived from the three curvature invariants $R, R_{ab}R^{ab}$ and $R_{abcd}R^{abcd}$. Although general relativity is well-supported by solar system tests, in the high curvature limit, such as on approach to an initial cosmological singularity, we expect quantum effects to become important and to cause deviations from the standard behaviour in general relativity. At high curvatures, anisotropies diverge faster than isotropies and will tend to dominate the cosmological behaviour at early times. Furthermore, it has previously been shown \cite{hervik} that anisotropies in these higher-order theories may display significantly different behaviour to that found in general relativity. Thus, in this work we have investigated the role of anisotropy on approach to the initial singularity.

In particular, we have found all Kasner-like solutions given by the Bianchi type I line element (\ref{kas}) for several wide classes of higher-order theories of gravity, and also all of the similar Bianchi I solutions which are described by the line element (\ref{ds}). Previously, Kasner-like solutions were known only for quadratic gravity \cite{deruelle} and higher-order Lagrangians which were powers of one of the curvature invariants $R, R_{ab}R^{ab}$ and $R_{abcd}R^{abcd}$. We have extended this to much more general Lagrangians including those of the form $f(R), f(R_{ab}R^{ab})$ and $f(R_{abcd}R^{abcd})$, and 
in the course of this investigation, we have also found additional solutions to the previously-studied power-law Lagrangians which were not found in \cite{clifton1, clifton2}. 

We have further widened this study to include also those Lagrangians which are homogeneous in time for the metrics (\ref{kas}), which may be written in the form (\ref{homogeneous}), using the model of Br\"{u}ning, Coule and Xu \cite{coule} as a particular example. Since the terms in the field equations must vanish at each order in time, then unless there exists a particular configuration of Kasner exponents such that the Lagrangian $f(X,Y,Z)$ is a constant, we expect that the solutions found in section \ref{sec:homogeneous} will be the \emph{only} Kasner-like solutions with real Kasner indices that are possible in higher-order metric theories of gravity derived from Lagrangians with the general form $f(R, R_{ab}R^{ab}, R_{abcd}R^{abcd})$. 

Although they are geometrically special, the Kasner-like solutions given by the Bianchi type I line element (\ref{kas}) provide us with a very useful insight into the dynamics of anisotropies, and also they give a good description of the evolution of more general anisotropic cosmological models over finite time intervals. This is of particular interest when considering the behaviour on approach to the initial singularity exhibited by the Bianchi type VIII and type IX (``Mixmaster'') cosmologies, which can be approximated by a sequence of different Kasner epochs. We have considered the properties of the solutions found in relation to the behaviour of these more general anisotropic cosmologies and found that in general it is model-dependent as to whether the universe will experience an infinite sequence of oscillations between Kasner regimes as the singularity is approached. A more detailed analysis is required to understand the extent of the validity of these vacuum  solutions in the presence of a non-comoving perfect fluid.

The conditions for the existence of de Sitter solutions in higher-order theories of gravity were already known \cite{clifton4}, and some examples of anistropically inflating solutions had previously been discovered for particular theories \cite{hervik}. However, we have extended this work and explicitly discovered the existence conditions for anistropically inflating solutions of the form (\ref{ds}) in all higher-order metric theories of gravity derived from Lagrangians with the general form $f(R, R_{ab}R^{ab}, R_{abcd}R^{abcd})$. For the simpler models in which the Lagrangian depends on just one of the curvature invariants $R, R_{ab}R^{ab}$, and $R_{abcd}R^{abcd}$, we found that some degree of fine-tuning is required in the Lagrangian for such solutions to exist, however this is not required in more general theories. These solutions explicitly demonstrate that the cosmic ``no-hair'' theorem of general relativity does not hold in general in higher-order theories.

\section*{Acknowledgements}The author wishes to thank John D. Barrow for useful discussions and acknowledges a STFC studentship.

%\begin{acknowledgements}The author wishes to thank John D. Barrow for useful discussions and acknowledges a STFC studentship.\end{acknowledgements}
\bibliography{bib2}

\end{document}